\documentclass[prd,twocolumn,preprintnumbers,nofootinbib]{revtex4-1}

\usepackage[utf8]{inputenc}

\usepackage{amsmath}
\usepackage{epsfig}
\usepackage{graphicx}
\usepackage{graphics}
\usepackage{color}
\usepackage{aas_macros}
\usepackage{bbm}
\usepackage{multirow}
\usepackage[normalem]{ulem}
\usepackage{booktabs}

\usepackage[breaklinks, plainpages=false, colorlinks=true, anchorcolor=cyan, linkcolor=red, citecolor=cyan, urlcolor=magenta, bookmarks=false]{hyperref}

\usepackage[caption=false]{subfig}
\usepackage[title]{appendix}
\usepackage{tikz,tikz-3dplot}

\DeclareMathOperator{\sinc}{sinc}

\begin{document}
\renewcommand{\thefigure}{\arabic{figure}}
\setcounter{figure}{0}

 \def\I{{\rm i}}
 \def\E{{\rm e}}
 \def\D{{\rm d}}

\bibliographystyle{apsrev}

\title{Detecting Gravitational Wave Bursts with LISA in the presence of Instrumental Glitches}

\author{Travis Robson}
\affiliation{eXtreme Gravity Institute, Department of Physics, Montana State University, Bozeman, Montana 59717, USA}

\author{Neil J. Cornish}
\affiliation{eXtreme Gravity Institute, Department of Physics, Montana State University, Bozeman, Montana 59717, USA}

\begin{abstract} 
The Laser Interferometer Space Antenna (LISA) will open a rich discovery space in the milli-Hertz gravitational wave band. In addition to the anticipated signals from many millions of binary systems, this band may contain new and previously un-imagined sources for which we currently have no models. To detect unmodeled and unexpected signals we need to be able to separate them from instrumental noise artifacts, or glitches. Glitches are a regular feature in the data from ground based laser interferometers, and they were also seen in data from the LISA Pathfinder mission. In contrast to the situation on ground, we will not have the luxury of having multiple independent detectors to help separate unmodeled signals from glitches, and new techniques have to be developed. Here we show that unmodeled gravitational wave bursts can be detected with LISA by leveraging the different way in which instrument glitches and gravitational wave bursts imprint themselves in the time-delay interferometery data channels. We show that for signals with periods longer than the light travel time between the spacecraft, the ``breathing mode'' or Sagnac data combination is key to detection. Conversely, for short period signals it is the time of arrival at each spacecraft that aids separation. We investigate the conditions under which we can distinguish the origin of signals and glitches consisting of a single sine-Gaussian wavelet and determine how well we can characterize the signal. We find that gravitational waves bursts can be unambiguously detected and characterized with just a single data channel (four functioning laser links), though the signal separation and parameter estimation improve significantly when all six laser links are operational.
\end{abstract}

\maketitle

\section{Introduction}\label{intro}

Gravitational wave astronomy has tremendous potential for discovery, as has been spectacularly demonstrated by the ground-based LIGO/Virgo observatories~\cite{PhysRevLett.116.061102, PhysRevLett.119.161101}. The signals that have been detected to-date have all been from binary systems, and are accurately modeled by theoretical templates. Going forward, it is hoped that entirely new classes of signals will be discovered, many of which we will not have templates for, either due to the difficulty in calculating the waveform (such as for supernovae), or from our ignorance about the existence of the source. Detecting signals of unknown morphology is challenging since the instruments themselves produce non-Gaussian transients, or glitches, that can be mistaken for signals of astrophysical origin. 

The Laser Interferometer Space Antenna (LISA)~\cite{2017arXiv170200786A}, like its ground based cousins, will very likely be afflicted by glitches. Glitches were seen in data from the LISA Pathfinder mission~\cite{PhysRevLett.116.231101, PhysRevLett.120.061101}, and it is hard to imagine that the will be absent from the more complex LISA measurement system. Characterizing these glitches and accurately estimating their waveforms, will be an important component of the LISA global data analysis program. Unlike the situation on ground, where the availability of multiple independent interferometers simplifies the task of separating glitches from signals, with LISA we will have a single instrument. Nor will we have any ``off-source'' data, free of loud gravitational wave signals, with which to perform a measurement of the instrument noise. With LISA  the signal and noise measurement must be done simultaneously~\cite{0264-9381-34-24-244002} as part of a global analysis.

Similar concerns led to the developments of burst and glitch characterization analyses for LIGO. One such analysis was the wavelet-based Bayesian algorithm BayesWave~\cite{0264-9381-32-13-135012}. This algorithm has played the key role for model-independent waveform reconstructions for most of the detected mergers seen by LIGO~\cite{PhysRevLett.116.061102}. Its broad capabilities were best demonstrated by the binary neutron star merger. BayesWave's ability to characterize a loud instrumental glitch, obscuring a large fraction of the all-important late inspiral, allowed for an accurate reconstruction of a the astrophysical signal with the glitch removed~\cite{Pankow:2018qpo}, so that other analyses could properly characterize the binary neutron star's physical parameters~\cite{PhysRevLett.119.161101}. For LISA, we wish to develop an algorithm to serve a similar purpose of analyzing glitches and bursts. Instrumental glitches in LISA studied here will fall into two categories: optical path and acceleration. Optical path glitches reflect non-Gaussian deviations in the optical path length of any of LISA's 6 laser links. Acceleration glitches result from disturbance to the acceleration of LISA's spacecraft. Laser phase noise glitches will be neglected in this study since they will be suppressed in the time delay interferometry (TDI) data channels~\cite{PhysRevD.65.082003}. Glitches can be represented through a superposition of sine-Gaussian wavelets in each component of the instrument. Gravitational wave bursts can similarly be represented by a superposition of wavelets. The signal is referenced to the solar system barycenter (SSB) and then projected onto LISA by computing the instrument response. As a first step, we consider signals and glitches that are described by a single wavelet and defer the generalization to multi-wavelet fits to future work.

To investigate our ability to characterize glitches and bursts consisting of the single wavelet, we will use Bayesian probability theory to calculate our degree of belief in the parameters which describe the injected signal, as quantified by the posterior distribution. The duration of these signals ranges from tens of seconds to roughly a day. Their duration, and frequency content in relation to the light travel time between the spacecraft will have important implications on our ability to characterize these signals and will also play a key role in our ability to distinguish whether the data contains a glitch (and which kind) or burst. Glitches will enter the time delay interferometry (TDI) data channels with time delays of the light travel time between spacecraft. This time is about 8.3 seconds for the nominal $L= 2.5$ Gm separation, which sets the LISA response transfer frequency  $f_* = c/(2\pi L)$ to be 19.1 mHz. Wavelets with frequencies below the transfer frequency will be harder to characterize and distinguish. An additional piece of the puzzle is which data channels the wavelets power crops up and in which proportion. Acceleration glitches enter the data stream by afflicting 2 different phase measurements while optical path glitches afflict only 1. Bursts, on the other hand, enter all phase measurements through time delays which depend on the various projected arm lengths depending on where the incident gravitational wave originates on the sky. While the power distribution is the most useful discriminant, the phasing becomes most important in the case of a malfunctioning LISA arm {\it i.e.} when when we would be left with only one data channel. In this study we will address these considerations and investigate what we can learn and what features are most informative.

This work is organized as follows: Section~\ref{sec:model} discusses the the waveforms for optical path and acceleration glitches and for gravitational wave bursts. Section~\ref{sec:bayes} reviews Bayesian inference and then describes the Markov Chain Monte Carlo algorithm we employ to carry out the parameter estimation and model selection analyses in this paper. Section~\ref{sec:PE} shows how well we can characterize glitch and burst parameters and recover the injected waveform. In section~\ref{sec:ModSel} we explore under what conditions we are able to distinguish a glitch and burst, and identify what features of the signal are most responsible for making the distinction. We end with a discussion of future work to be carried out in section~\ref{sec:discuss}. Note that we work in units where $G=c=1$.


\section{Glitch and Burst Models}\label{sec:model}
The LISA constellation consists of 3 spacecraft in the shape of a quasi-equilateral triangle trailing behind Earth. The spacecraft have a total of 6 laser links, 2 for each arm. Each laser get is phase measured by phasemeters onboard of the LISA spacecraft. A photon sent from the laser situated on spacecraft $i$, pointing towards spacecraft $j$, is emitted at time $t - L_{ij}$, where $L_{ij}$ is arm length connecting spacecraft $i$ and $j$. The phase of this photon is measured at time $t$ by the phasemeter on spacecraft $j$. This phase measurement can be approximated as~\cite{PhysRevD.69.082003}

\begin{align}
\Phi_{ij}(t) &= C_{i}(t-\mathrm{L}_{ij}) - C_{j}(t) + \psi_{ij}(t) + n^{\mathrm{o}}_{ij}(t) \nonumber \\ 
&~~~-\hat{\textbf{r}}_{ij}(t)\cdot\left(\textbf{n}^{\mathrm{a}}_{ji}(t-L_{ij}) -  \textbf{n}^{\mathrm{a}}_{ij}(t) \right)\,\,.
\label{equ:phase}
\end{align}
The noise in the laser phase itself is described by the terms $C_{i}$. The term $\psi_{ij}$ describes the phase shift induced by the presence of gravitational waves. The term $n^{\mathrm{o}}_{ij}$ represents the contribution from the optical bench on spacecraft $j$ that receives light from spacecraft $i$. The last term represents the contribution to the phase measurement incurred by the acceleration noise of the spacecrafts. Note that in this simplified model, where we are neglecting higher order features of LISA's motion such as the flexing of the arms, the only component of the acceleration that is relevant is the differential acceleration along the line $\hat{\textbf{r}}_{ij}$ connecting the center of mass of the two spacecraft.

These laser phase measurements are expected to be dominated by laser phase noise. Current estimates indicate that the laser phase noise will be roughly 10 orders of magnitude greater that the changes in phase induced by the gravitational waves of interest~\cite{PhysRevD.65.082003}. The phase noise can be canceled using time delay interferometry (TDI). The TDI data combinations synthesize light paths of equal length by adding together phase measurements with time delays given by multiples of the instantaneous light travel times. This superposition cancels the laser phase noise. When higher order corrections to the spacecraft motion are taken into account the superposition of time delayed phase measurements become more complicated. Here we use the simpler first generation TDI data combinations. Three Michelson-like TDI channels can be formed from the signals extracted at each vertex of the observatory. These are denoted as $X$, $Y$, and $Z$. The $X$ TDI channels is constructed as follows:
\begin{align}
X(t) &= \Phi_{12}(t-3\mathrm{L}) - \Phi_{13}(t - 3\mathrm{L}) + \Phi_{21}(t-2\mathrm{L}) \nonumber \\
&~~~ - \Phi_{31}(t-2\mathrm{L}) + \Phi_{13}(t-\mathrm{L}) - \Phi_{12}(t-\mathrm{L}) \nonumber \\
&~~~+ \Phi_{31}(t) - \Phi_{21}(t) \,\,,
\label{equ:X}
\end{align}
where we have assumed that the LISA arm lengths are of constant length, i.e. $L_{ij}(t) = L$. The $Y$ and $Z$ channels are constructed through a cyclic permutation of the spacecraft labels---e.g. $1\rightarrow2$, $2\rightarrow3$, and $3\rightarrow1$ to construct the $Y$ TDI channel. It is often convenient to work with following linear combinations of the $X,Y,Z$ channels:
\begin{subequations}
	\begin{align}
	A &= \frac{1}{3}(2X-Y-Z) \\
	E &= \frac{1}{\sqrt{3}}(Z-Y) \\
	T &= \frac{1}{3}(X+Y+Z) \,\,.
	\label{equ:AET}
	\end{align}
\end{subequations}

Below the transfer frequency, the $A$ and $E$ channels synthesize two right angle interferometers with a relative orientation of $45^\circ$, and provide instantaneous measures of the plus and cross polarization states of a gravitational wave. At these frequencies the $T$ channel is mostly sensitive to the scalar breathing mode polarization state, which is absent in Einstein gravity, and thus provides a null channel that is useful for measuring a particular combination of the noise contributions. When the noise levels are equal on each spacecraft, the cross-spectral density of the noise in the $A,E,T$ channels vanish~\cite{PhysRevD.82.022002,PhysRevD.89.022001}.

An arbitrary signal seen in the TDI data channels may be reconstructed by a superposition of sine-Gaussian wavelets. In this study we use Gabor-Morlet wavelets. In the time-domain they are given by 
\begin{equation}
\Psi = A e^{-(t-t_{0})^{2}/\tau^{2}}\cos\left[2\pi f_{0}(t-t_{0}) + \phi_{0}\right] \,\,,
\label{equ:wavelet}
\end{equation}
where $A$ is the wavelet amplitude, $t_{0}$ and $f_{0}$ are the central time and frequency, the wavelet time scale is $\tau$---related to the wavelet quality factor $Q$ through $\tau~=~Q/2\pi f_{0}$---and $\phi_{0}$ is the initial phase. Occasionally we will use the variable $\bar{\phi} = \phi_{0} - 2\pi f_{0} t_{0}$. The Fourier transform of the Gabor-Morlet wavelet is 

\begin{align}
\tilde{\Psi} &= \frac{\sqrt{\pi}\tau A}{2}e^{-i(2\pi f t_{0} + \phi_{0})} \nonumber \\
&~~~~~ \times\left[e^{-\left(\pi\tau(f+f_{0})\right)^{2}} + e^{2i\phi_{0}}e^{-\left(\pi\tau(f-f_{0})\right)^{2}} \right] \,\,.
\label{equ:FTwavelet}
\end{align}
In the Fourier domain we see in the large quality factor $Q$, or equivalently large $\tau$ regime, the second term in eqn.~(\ref{equ:FTwavelet}) is dominant. Ignoring the sub-dominant term, we can estimate the signal-to-noise ratio (SNR) in the case of white noise as
\begin{align}
\rho^{2} \approx& \frac{4}{S_{n}(f_{0})}\int_{0}^{\infty} \left(\frac{\sqrt{\pi}\tau A}{2}\right)^{2}e^{-2\left(\pi\tau(f-f_{0})\right)^{2}} df \nonumber \\
\approx & \sqrt{\frac{\pi}{2}}\frac{A^{2} \tau}{S_{n}(f_{0})} \,\,,
\label{eq:snrEST}
\end{align}
where $S_{n}(f_{0})$ is an appropriate noise power spectral density which has been assumed constant such that we may approximate the integral. This result will become useful later when we wish to estimate a reasonable bandwidth in the frequency domain to calculate these signals over.


\subsection{Instrumental Glitches}\label{glitch}

To model instrumental glitches we inject a Gabor-Morlet wavelet into the appropriate term in eqn. (\ref{equ:phase}). For example, a glitch in the optical path length pointing from spacecraft 1 to 2 is modeled as $\Phi_{12}(t) = n^{\mathrm{o}}_{12}(t)  = \Psi(t)$. We will label such a glitch as $\Phi_{12}^{\mathrm{op}}$. For an acceleration glitch associated with the proof mass on spacecraft 2 that is referenced against spacecraft 1 will appear in to phase measurements: $\Phi_{12}(t)  = -\Psi(t)$ and $\Phi_{21}(t)  = \Psi(t-L)$. This acceleration glitch will be denoted as $\Phi_{12}^{\mathrm{ac}}$. Laser phase glitches are neglected in this work since the TDI channels are constructed such that laser phase noise is canceled.

The $X$, $Y$, and $Z$ TDI channels can be constructed for both optical path and accelerations glitches analytically in the frequency domain. For the optical path glitch $\Phi_{12}^{\mathrm{op}}$ the response is
\begin{subequations}
\begin{align}
\tilde{X} &= 2i\tilde{\Psi}e^{-2if/f_{*}}\sin\frac{f}{f_{*}}\\
\tilde{Y} &= -2i\tilde{\Psi}e^{-if/f_{*}}\sin\frac{f}{f_{*}} \\
\tilde{Z} &= 0 \, .
\end{align}
\end{subequations}
Note that there is no response in the $Z$ channel. The factor of $\sin f/f_{*}$ is due to differencing the disturbance by the time delay. The only other optical path glitch that has no response in the $Z$ channel is
 $\Phi_{21}^{\mathrm{op}}$, which produces the response
\begin{subequations}
\begin{align}
\tilde{X} &= 2i\tilde{\Psi}e^{-if/f_{*}}\sin\frac{f}{f_{*}} \,\,,\\
\tilde{Y} &= -2i\tilde{\Psi}e^{-2if/f_{*}}\sin\frac{f}{f_{*}} \\
\tilde{Z} &= 0 \, .
\end{align}
\end{subequations}
We can already glean insight into how optical patch glitches can be identified. When all 6 laser links are functioning, none of the optical path glitches can be made to look like the other. For example, suppose we try to match the $X$ channel response of $\Phi_{12}^{\mathrm{op}}$ to that of $\Phi_{21}^{\mathrm{op}}$. This would require a time shift of $t+L$ i.e. a factor of $e^{if/f_{*}}$ in the frequency domain. This time shift will of course shift the $Y$ response in the opposite desired direction in time. We cannot find a transformation of wavelet parameters such that any optical path glitch looks like another when all 6 laser links functioning. If we are unfortunate enough to have only 2 functioning arms, we will be at a loss when attempting to distinguish these two glitches. That is if we have only the $X$ channel, we will not be able to distinguish $\Phi_{12}^{\mathrm{op}}$ from a time shifted $\Phi_{21}^{\mathrm{op}}$.

We must also contend with acceleration glitches. The acceleration glitch $\Phi^{\mathrm{ac}}_{12}$ has the TDI response
\begin{equation}
\tilde{Y} = 4 \tilde{\Psi} e^{-2if/f_{*}}\sin^{2}\frac{f}{f_{*}} \,\,.
\end{equation}
Where both the $X$ and $Z$ channel are null. All acceleration glitches have a response in only 1 of the $X$, $Y$, and $Z$ data channels. Acceleration glitches also have an additional suppression from the extra factor the transfer function $\sin f/f_{*}$. This is due to the acceleration glitch appearing in two phase measurements separated by the light travel time between spacecraft. With acceleration glitches however, we are unable to unambiguously determine their origin even when all 6 laser links are functioning. There are perfect degeneracies between pairs of acceleration glitches. For example, the response to the acceleration glitch $\Phi^{\mathrm{ac}}_{32}$ is
\begin{equation}
\tilde{Y} = -4 \tilde{\Psi} e^{-2if/f_{*}}\sin^{2}\frac{f}{f_{*}} \,\,.
\end{equation}
has precisely the same form as the $\Phi^{\mathrm{ac}}_{12}$ except a shift in its initial phase (by $\pi$). In the scenario that we lose one arm of the constellation we will be no worse off with respect to distinguishing acceleration glitches. The response to glitches in other components are shown in table~\ref{tbl:TDI}.

\begin{table*}[t]
\centering
\renewcommand\arraystretch{1.5}
\begin{tabular}{ |c|c|c|c|c| } 
\hline
& $\tilde{X}$ & $\tilde{Y}$ & $\tilde{Z}$ \\
\hline
$\Phi^{\mathrm{op}}_{12}$ & $2i\tilde{\Psi}e^{-2i f/f_{*}}\sin \left(f/f_{*}\right)$ & $-2i\tilde{\Psi}e^{-i f/f_{*}}\sin\left(f/f_{*}\right)$ & 0 \\ 
$\Phi^{\mathrm{op}}_{21}$ & $2i\tilde{\Psi}e^{-i f/f_{*}}\sin\left(f/f_{*}\right)$ & $-2i\tilde{\Psi}e^{-2i f/f_{*}}\sin\left(f/f_{*}\right)$ & 0 \\ 
$\Phi^{\mathrm{op}}_{13}$ & 0 & $2i\tilde{\Psi}e^{-i f/f_{*}}\sin\left(f/f_{*}\right)$ & $-2i\tilde{\Psi}e^{-2i f/f_{*}}\sin\left(f/f_{*}\right)$ \\ 
$\Phi^{\mathrm{op}}_{31}$ & 0 & $-2i\tilde{\Psi}e^{-i f/f_{*}}\sin\left(f/f_{*}\right)$ & $2i\tilde{\Psi}e^{-2i f/f_{*}}\sin\left(f/f_{*}\right)$ \\ 
$\Phi^{\mathrm{op}}_{23}$ & $2i\tilde{\Psi}e^{-i f/f_{*}}\sin\left(f/f_{*}\right)$ & 0 & $-2i\tilde{\Psi}e^{-2i f/f_{*}}\sin\left(f/f_{*}\right)$ \\ 
$\Phi^{\mathrm{op}}_{32}$ & $2i\tilde{\Psi}e^{-i f/f_{*}}\left(f/f_{*}\right)$ & 0 & $-2i\tilde{\Psi}e^{-2i f/f_{*}}\sin\left(f/f_{*}\right)$ \\ 
\hline
$\Phi^{\mathrm{ac}}_{12}$ & 0 & $4\tilde{\Psi}e^{-2i f/f_{*}}\sin^{2}\left(f/f_{*}\right)$ & 0 \\ 
$\Phi^{\mathrm{ac}}_{21}$ & $-4\tilde{\Psi}e^{-2i f/f_{*}}\sin^{2}\left(f/f_{*}\right)$ & 0 & 0 \\ 
$\Phi^{\mathrm{ac}}_{13}$ & 0 & 0 & $-4\tilde{\Psi}e^{-2i f/f_{*}}\sin^{2}\left(f/f_{*}\right)$ \\ 
$\Phi^{\mathrm{ac}}_{31}$ & $4\tilde{\Psi}e^{-2i f/f_{*}}\sin^{2}\left(f/f_{*}\right)$ & 0 & 0 \\ 
$\Phi^{\mathrm{ac}}_{23}$ & 0 & 0 & $4\tilde{\Psi}e^{-2i f/f_{*}}\sin^{2}\left(f/f_{*}\right)$ \\ 
$\Phi^{\mathrm{ac}}_{32}$ & 0 & $-4\tilde{\Psi}e^{-2i f/f_{*}}\sin^{2}\left(f/f_{*}\right)$ & 0 \\ 
\hline
\end{tabular}
\caption{\label{tbl:TDI} This table contains the analytic first generation TDI variables for optical path and acceleration glitches. Note that optical path glitches occupy 2 of the $XYZ$ TDI channels while the acceleration glitches only occupy 1. Acceleration  glitches pick up an additional factor of the transfer function $\sin f/f_{*}$. A change in wavelet parameters, specifically the initial phase, leads to a perfect degeneracy between pairs of acceleration glitches when all three TDI channels are functioning. This is not the case for optical path glitches.}
\end{table*}

When generating these waveforms we wish economically sample an appropriate bandwidth. The signal-to-noise ratio for an optical path glitch, given one data channel, can be estimated as

\begin{equation}
\rho_{\mathrm{est}}^{2} = \frac{\sqrt{\pi/2} A^{2} \tau}{S_{X,M}(f_{0})} \,\,,
\end{equation}
in the large $\tau$ limit obtained from eqn.~(\ref{eq:snrEST}). $S_{X,M}$ is the Michelson-equivalent power spectral density defined as $S_{X,M} = S_{X}/4 \sin^{2}(f/f_{*})$. Similarly, the SNR for an acceleration glitch can be estimated as 

\begin{equation}
\rho_{\mathrm{est}}^{2} = \frac{4 \sqrt{\pi/2} A^{2} \tau \sin^{2}(f_{0}/f_{*})}{S_{X,M}(f_{0})} \,\,.
\end{equation}
A bandwidth of $\Delta f = 4 (\rho_{\mathrm{est}}/5)^{2}/\tau$ was used to capture in excess of $99.9\%$ of the SNR in each glitch.

\begin{figure*}[!htb]
\centering
\includegraphics[width=1.0\textwidth]{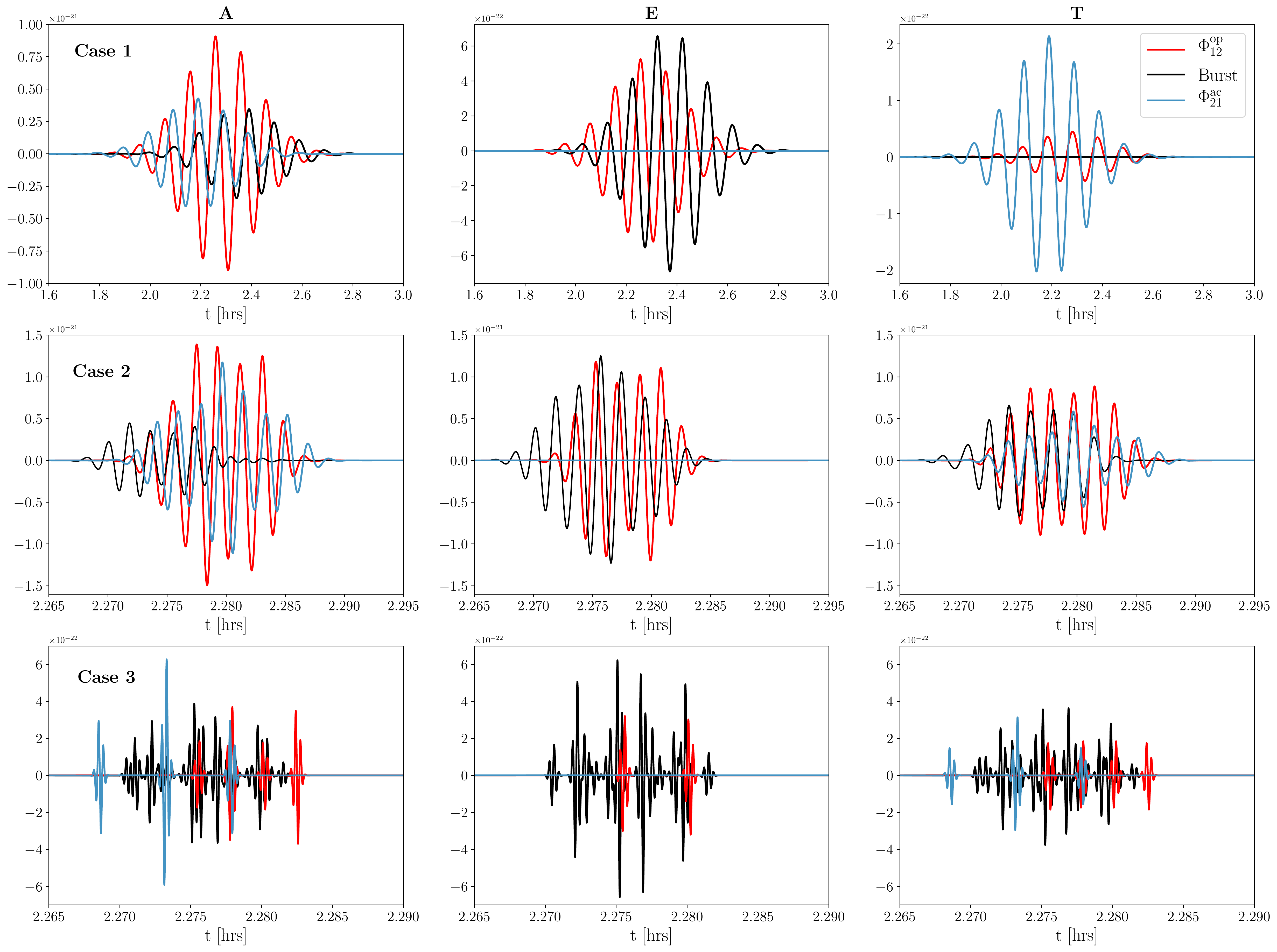} 
\caption{\label{fig:ThreeRegimes} This figure displays the $AET$ TDI channel responses for various glitches ($\Phi_{12}^{\mathrm{op}}$ in red, and $\Phi_{21}^{\mathrm{ac}}$ in blue) and gravitational wave bursts (in black). The top row shows wavelets with durations that are much longer than the light travel time between LISA spacecraft. The middle row shows wavelets with duration that is comparable to the light travel time. The bottom row shows wavelet with duration less than the light travel time, which leads to a clean separation of the glitch wavelets in the TDI channels. Note that the glitch wavelets only appear in a subset of the TDI channels.}
\end{figure*}

In addition to distributing power in different TDI channels, glitches in different components produce different phasing in the response. The phasing information depends critically on the frequency of the glitch, $f_0$ and the duration of the glitch $\tau$. Higher frequency glitches get heavily modulated by the transfer functions making it easier to determine their origin. In Figure~\ref{fig:ThreeRegimes} the $AET$ TDI channels for optical path glitches $\Phi_{12}^{\mathrm{op}}$ are displayed in red and acceleration glitches $\Phi_{21}^{\mathrm{ac}}$ in shown blue.
The amplitudes of the optical and acceleration glitches were chosen for ease of comparison, while maintaining the correct relative amplitudes in the different TDI channels.  The top row (case 1) displays glitches with the parameters $\tau = 0.2$~hours, and $f_{0} = 2/\tau$ i.e. $2.7$~mHz, placing this glitch well below the transfer frequency. These parameters give the glitch a quality factor of 12.6. Since the wavelet has a low frequency, its amplitude does not change substantially in a light travel time. This means that the construction of the TDI channels acts like a derivative of the input. In the middle row (case 2) the parameters of the wavelet are $\tau = L = 8.33$~seconds and $f_{0} = 1.3/\tau = 156$~mHz ($Q\sim 8$). This wavelet has a temporal extent comparable to that of the light travel time between spacecraft. This results in a waveform that is the superposition of two wavelets with a small time shift between them. Lastly, in the bottom row (cases 3) we see a wavelet of $\tau=1$~second and $f_{0} = 800$~mHz i.e. $Q = 5.1$. Here the frequency of the signal is substantially larger than the transfer frequency and the duration of the signal in time is much less than the light travel time, leading to a clean separation of the wavelets in the TDI channels. Note that in the low frequency regime the optical path glitch has a suppressed output in the $T$ channel. We also see that the $E$ channel response to the acceleration glitch is totally suppressed. This is because there is no $Z$ or $Y$ response for this specific acceleration glitch and the $E$ channel has no $X$ channel dependence.


\subsection{Gravitational Wave Bursts}\label{burst}

The optical path length change due to a  gravitational wave signal in the laser link connecting the $i^{\mathrm{th}}$ and $j^{\mathrm{th}}$ spacecraft is given by
\begin{equation}
\delta \ell_{ij}(t) = \textbf{D}_{ij}:\int_{\xi_{i}}^{\xi_{j}}\textbf{h}(t) dt\,\,,
\label{equ:dl}
\end{equation}
where the colon denotes full contraction between the tensors, i.e. $\textbf{A}:\textbf{B} = A^{jk}B_{jk}$. The time $t$ is Solar System Barycenter (SSB) time, and $\xi_{i} = t_{i} - \hat{\textbf{k}}\cdot\textbf{x}_{i}$ is the wave variable defining surfaces of constant phase for the gravitational wave. The position of the $i^{\mathrm{th}}$ spacecraft is $\textbf{x}_{i}$ and $t_{i}$ is the time of emission of the laser photon from spacecraft $i$ and $t_{j}$ is the time of reception of the laser photon at spacecraft $j$. The detector tensor $\textbf{D}$ is given by
\begin{equation}
\textbf{D} = \frac{1}{2}\frac{\hat{\textbf{r}}_{ij}\otimes\hat{\textbf{r}}_{ij}}{1-\hat{\textbf{k}}\cdot\hat{\textbf{r}}_{ij}} \,\,,
\label{equ:det_tensor}
\end{equation}
where $\hat{\textbf{k}} = -(\sin\theta \cos\phi, \sin\theta \sin\phi, \cos\theta)$ is the gravitational wave propagation direction---$\theta$ and $\phi$ designate the source's position in spherical polar coordinates in the SSB frame. The quantity $\hat{\textbf{r}}_{ij}$ is the unit-separation vector between the LISA spacecraft pointing from spacecraft $i$ to spacecraft $j$. In this study the LISA orbits are kept to leading order in eccentricity thereby fixing LISA arm length to be constant~\cite{PhysRevD.69.082003} $\mathrm{L} = |\textbf{r}_{ij}|$ for all $i,j$ combinations.

The gravitational wave tensor $\textbf{h}$ is given by

\begin{equation}
\textbf{h} = h_{+}(t) \textbf{e}'_{+}(\psi,\theta,\phi) + h_{\times}\textbf{e}'_{\times}(\psi,\theta,\phi)   \,\,,
\label{equ:gw_tensor}
\end{equation}
where $\textbf{e}'_{+,\times}$ are the polarization tensors $\textbf{e}_{+,\times}$ rotated by the polarization angle $\psi \in \left[0, \pi\right]$. In this work we assume that the gravitational waves are elliptically polarized such that, in the frequency domain, $\tilde{h}_{\times} = i \epsilon \tilde{h}_{+}$ parameterized by the ellipticity $\epsilon \in \left[0,1\right]$. We model the integrated gravitational wave polarizations as Gabor-Morlet wavelets such that $\int^{t}h_{+}(t')dt' = L \Psi(t)$. We may approximate the detector as static for the duration of a wavelet, since corrections would be on the order of $\tau/1 \mathrm{yr}$. This means we may safely evaluate all terms associated with the position of the detector at the central time of the wavelet $t_{0}$ and assume the value is constant. The response to the wavelet in the frequency domain is then

\begin{align}
\tilde{y}_{ij}&= \frac{f}{f_{*}}  \left(F^{+'}_{ij} + i \epsilon F^{\times'}_{ij}\right)  \mathcal{T}_{ij}(f; \hat{\textbf{k}}) \tilde{\Psi}(f)e^{-2\pi i f \hat{\textbf{k}} \cdot \textbf{x}_{i}(t_{0})}
\label{eq:gwRep}
\end{align}
where $y_{ij} = \delta \ell_{ij}/2L$ and $F^{+,\times}_{ij} =  \left[\hat{\textbf{r}}_{ij}\otimes \hat{\textbf{r}}_{ij}\right]: \textbf{e}_{+,\times}$. $\mathcal{T}_{ij}$, the transfer function, is given by
\begin{align}
\mathcal{T}_{ij} =&  \frac{1}{4} e^{i\left( \frac{f}{2 f_{*}}(1 - \hat{\textbf{k}}\cdot\hat{\textbf{r}}_{ij})\right)} \sinc \left( \frac{f}{2 f_{*}}(1 - \hat{\textbf{k}}\cdot\hat{\textbf{r}}_{ij})\right)\,\,.
\label{eq:transfer}
\end{align}
The wavelet has its central time shifted by the light travel time between the SSB origin and spacecraft $i$ through the phase factor $e^{2\pi i f \hat{\textbf{k}} \cdot \textbf{x}_{i}(t_{0})}$ present in eqn.~(\ref{eq:gwRep}). The ellipticity $\epsilon$ and polarization angle $\psi$ simply modulate the amplitude of the response. We see that the sky angles modulate the amplitude too, but also enter into the phasing. As opposed to instrumental glitches, gravitational wave bursts will induce responses in all TDI channels. It is important to note though that for frequencies below the transfer frequency  $f_{*}$,  the gravitational wave response in the $T$ channel is heavily suppressed~\cite{PhysRevD.82.022002}. This can be seen in the $T$ channel response for case 1 in Figure~\ref{fig:ThreeRegimes} .  The signal in each panel of Figure~\ref{fig:ThreeRegimes} represents a gravitational wave burst. The sky angles were chosen such that $\cos\theta = 0.23$ and $\phi=2.31$. The polarization angle was $0.45$ and the ellipticity $0.5$. The wavelet parameters are precisely the same as those for the glitches in the panel the burst shares (give or take an amplitude factor or time shift for the sake of easy comparison). Note that for case 1 the signal response is distinctly different than for the glitch. Recall that the glitches in this case were cleanly separated. This is because glitches enter the data stream with time delays equal to the light travel time between spacecraft, which is longer than their extent in time. Gravitational waves enter the data stream with time delays equal to the projected arm lengths. This can lead to foreshortened arms allowing for some overlap between the wavelets as see in case 3. 

\section{Bayesian Inference }\label{sec:bayes}

With the glitch and burst models established we now turn to the methods used to infer the properties of the gravitational wave signals and instrument glitches and develop probability distributions for the parameters of the models. These probabilities are quantified by the posterior distribution $p(\vec{\lambda}_{\tiny \mathcal{M}}|\textbf{s}, \mathcal{M})$ which reflects our belief about a given set of parameters $\vec{\lambda}_{\tiny \mathcal{M}}$ which specify model $\mathcal{M}$ given data $\textbf{s}$. The posterior distribution is obtained via Baye's theorem:

\begin{equation}
p(\vec{\lambda}_{\tiny \mathcal{M}}|\textbf{s}, \mathcal{M}) = \frac{p(\textbf{s}|\vec{\lambda}_{\tiny \mathcal{M}}, \mathcal{M})p(\vec{\lambda}_{\tiny \mathcal{M}}|\mathcal{M})}{p(\textbf{s}|\mathcal{M})} \,\,,
\end{equation}
where $p(\vec{\lambda}_{\tiny \mathcal{M}}|\mathcal{M})$ is the prior distribution for the parameters $\vec{\lambda}_{\tiny \mathcal{M}}$, $p(\textbf{s}|\vec{\lambda}_{\tiny \mathcal{M}}, \mathcal{M})$ the likelihood of the data given the parameters, and $p(\textbf{s}|\mathcal{M})$ is the evidence for the model $\mathcal{M}$. Along with the assumptions we have already made in the construction of the TDI channels, we further assume that, aside from the glitches modeled here, the noise is stationary and Gaussian. The likelihood function for the data then takes the form

\begin{equation}
p(\textbf{s}|\vec{\lambda}, \mathcal{M}) \propto \exp\left[ -\frac{1}{2} \sum_{I} \left(\textbf{s}_{I}-\textbf{h}_{I}(\vec{\lambda})|\textbf{s}_{I}-\textbf{h}_{I}(\vec{\lambda})\right) \right] \,\,, 
\end{equation}
where the $\mathcal{M}$ subscript on the parameters has been dropped for simplicity. The sum is over TDI data streams $I = \lbrace A, E, T\rbrace$ (or just $I = \lbrace X \rbrace$ for some of our investigations). The noise-weighted inner product is defined as

\begin{equation}
(\textbf{a}_{I}|\textbf{b}_{I}) = 4\mathcal{R}\int_{0}^{\infty} \frac{\tilde{a}_{I}(f)\tilde{b}_{I}^{*}(f)}{S_{n,I}(f)}df \,\,.
\label{equ:nwip}
\end{equation}
The noise strain spectral density in these data channels are given by

\begin{widetext}
	\begin{subequations}
		\begin{align}
		S_{AE} &= \frac{16}{3}\sin^{2}\frac{f}{f_{*}}\left[\left(2+\cos\frac{f}{f_{*}}\right)P_{\mathrm{OMS}} + 2\left(3 + 2\cos\frac{f}{f_{*}} + \cos \frac{2f}{f_{*}}\right)\frac{P_{\mathrm{acc}}}{(2\pi f)^{4}}\right]\frac{1}{(2\mathrm{L})^{2}} \\
		S_{T} &= \frac{16}{3}\sin^{2}\frac{f}{f_{*}}\left[\left(1-\cos\frac{f}{f_{*}}\right)P_{\mathrm{OMS}} + \left(3-4\cos \frac{f}{f_{*}} + \cos\frac{2f}{f_{*}}\right)\frac{P_{\mathrm{acc}}}{(2\pi f)^{4}} \right]\frac{1}{(2\mathrm{L})^{2}} \\
		S_{X} &= 4 \sin^{2}\frac{f}{f_{*}}\left[4 P_{\mathrm{OMS}} + 8 \left(1+ \cos^{2} \frac{f}{f_{*}}\right)\frac{P_{\mathrm{acc}}}{(2\pi f)^{4}} \right]\frac{1}{(2\mathrm{L})^{2}}  \, .
		\end{align}
	\end{subequations}
\end{widetext}
The noise in the $A$ and $E$ channels, $S_{AE}$, is the same and the noise in the $T$ channel is $S_{T}$. The single-link optical metrology noise power $P_{\mathrm{OMS}}$ and single test-mass acceleration noise power $P_{\mathrm{acc}}$ are quoted in~\cite{2018arXiv180301944C}. Another contribution to the measured noise comes from millions of unresolved galactic binaries~\cite{1742-6596-840-1-012024} emitting gravitational waves to which LISA is sensitive. Estimates of the unresolved galactic binary confusion noise for various observation periods can also be found in the same reference.


\subsection{Maximization over nuisance parameters}

In a fully Bayesian analysis we would compute the joint posterior distributions all parameters in the model. To simplify the analysis and achieve more rapid convergence, we chose to eliminate certain nuisance parameters by analytically maximizing the likelihood with respect to these parameters using the F-statistic approach (we could have analytically {\em marginalized} over the nuisance parameters instead~\cite{Thrane:2018qnx}, but it is much faster and simpler to maximize). The F-statistic~\cite{PhysRevD.72.043005} provides a way to maximize the likelihood over the extrinsic parameters---$A$, $\phi_{0}$ for a glitch, and $A$, $\phi_{0}$, $\psi$, $\epsilon$ for a burst. Through the use of several filters, constructed from the burst or glitch wavelet with specific choices of extrinsic parameters, one may construct the maximize likelihood. To understand how to construct the F-statistic it is useful to consider the burst model in the large $\tau$ and low frequency limit

\begin{align}
y_{ij} =&-\frac{f_{0}}{4 f_{*}} \left[ F_{ij}^{+'} \sin\left(2\pi f_{0} t_{i} + \bar{\phi} \right)  \right.\nonumber \\
&\left.  + \epsilon F_{ij}^{\times '} \cos\left(2\pi f_{0} t_{i} + \bar{\phi} \right)  \right] \,\,,
\end{align}
where $t_{i} = t - \hat{\textbf{k}}\cdot\textbf{x}_{i}$. This signal may be deconstructed into four terms which consist of a constant amplitude dependent on extrinsic parameters multiplying a time-dependent factor, additionally  dependent on the intrinsic parameters ($f_{0}$, $\tau$, $t_{0}$, $\theta$, $\phi$)

\begin{equation}
y_{ij} = \sum_{k} a_{k} A^{k}(t) \,\,.
\end{equation}
The four filters $A^{k}(t)$

\begin{subequations}
\begin{align}
A^{1} &= -\frac{f_{0}}{4f_{*}}F_{ij}^{+} \sin\left(2\pi f_{0} t_{i} \right) \\
A^{2} &= -\frac{f_{0}}{4f_{*}}F_{ij}^{\times} \sin\left(2\pi f_{0} t_{i}  \right) \\
A^{3} &= -\frac{f_{0}}{4f_{*}}F_{ij}^{+}\cos\left(2\pi f_{0} t_{i}  \right) \\
A^{4} &=- \frac{f_{0}}{4f_{*}}F_{ij}^{\times} \cos\left(2\pi f_{0} t_{i}  \right)
\end{align}\label{eq:filters}
\end{subequations}
may be constructed by inserting the following the extrinsic parameters, as described by table~\ref{tbl:Fstat}, into the burst waveform generator.
\begin{table}[t]
	\renewcommand\arraystretch{1.4}
	\centering
	\begin{tabular}{ |p{10mm}|p{10mm}|p{10mm}|p{10mm}|p{10mm}| } 
		\hline
		Filter & $A$ & $\bar{\phi}$ & $ \psi$ & $\epsilon$ \\
		\hline
		$A^{1}$ & 1 & 0 & 0 & 0 \\ 
		$A^{1}$ & 1 &0 & $-\pi/4$ & 0 \\
		$A^{3}$ & 1 & $\pi/2$ & 0  & 0\\ 
		$A^{4}$ & 1 &$\pi/2$  & $-\pi/4$ & 0 \\ 
		\hline
	\end{tabular}\label{tbl:Fstat}
	\caption{Plugging these parameters into the gravitational wave burst waveform generator will construct the filters eqns.~(\ref{eq:filters}). The resulting filters can then be used to maximize the likelihood over the extrinsic parameters.}
\end{table}
The glitch F-statistic filters can be constructed by the parameter choices: 1) $A=1, \bar{\phi} = 0$ and 2) $A=1, \bar{\phi} = -\pi/4$. The extrinsic parameter coefficients are 

\begin{subequations}
\begin{align}
a_{1} = & A\left( \cos 2\psi \cos \bar{\phi} - \epsilon \sin 2\psi \sin\bar{\phi} \right) \\
a_{2} = & A\left( -\sin 2\psi \cos \bar{\phi} - \epsilon \cos 2\psi \sin\bar{\phi} \right) \\
a_{3} = & A\left( \cos 2\psi \sin \bar{\phi} + \epsilon \sin 2\psi \cos\bar{\phi} \right) \\
a_{4} = & A\left( -\sin 2\psi \sin \bar{\phi} + \epsilon \cos 2\psi \cos\bar{\phi} \right)  \,\,.
\end{align}
\end{subequations}
The noise-weighted inner product of these filters with the data $N^{k} = (\textbf{s}|\textbf{A}^{k})$ can be used to construct the maximized relative likelihood

\begin{equation}
\mathcal{F} = \log \mathcal{L} = \frac{1}{2}\left(M^{-1}\right)_{mn} N^{m} N^{n} \,\,.
\end{equation}
The value $\mathcal{L}$ is the relative likelihood, i.e. the ratio between the likelihood assuming $\textbf{h}$ contains a burst and the likelihood assuming there is no such signal, i.e. $\textbf{h} = 0$, such that

\begin{equation}
\log \mathcal{L} = (\textbf{s}|\textbf{h}) - \frac{1}{2}(\textbf{h}|\textbf{h}) \,\,.
\end{equation}
The results hold for summing over multiple data channels such as when we use the $AET$ TDI channels. The matrix $M^{m n} = (\textbf{A}^{m}|\textbf{A}^{n})$ is simply the inner product matrix of the filters. Although, in this study we do not make use of the extrinsic parameters which maximize the likelihood it may prove useful in a future to study to be able to calculate them. Inverting the equations for the filter returns the extremized extrinsic parameters

\begin{subequations}
\begin{align}
	A =& \sqrt{\frac{1}{2}\left(s + \sqrt{pq}\right)} \\
	\epsilon =& \frac{s - \sqrt{pq}}{2\left (a_{1}a_{4} - a_{2}a_{3}\right) } \\
	\tan(2\psi) =& \frac{a_{1}^{3} + 2 a_{2} a_{3}a_{4} + a_{1}(a_{2}^{2} +a_{3}^{2} - a_{4}^{2} + \sqrt{pq})}{a_{1}^{2}a_{2} + 2 a_{1} a_{3}a_{4} + a_{2}(a_{2}^{2} -a_{3}^{2} + a_{4}^{2} + \sqrt{pq})}  \\
	\tan\bar{\phi} =&  \frac{a_{1}^{2} + a_{2}^{2} - a_{3}^{2} - a_{4}^{2} + \sqrt{pq}}{-2(a_{1}a_{3} + a_{2}a_{4})}
\end{align}
\end{subequations}
where $s = a_{1}^{2} + a_{2}^{2} + a_{3}^{2} + a_{4}^{2}$, $p=(a_{2}+a_{3})^{2} + (a_{1}-a_{4})^{2}$, and $q=(a_{2}-a_{3})^{2} + (a_{1}+a_{4})^{2}$. For glitches the amplitude and phase can be extracted via

\begin{subequations}
\begin{align}
A =& \sqrt{a_{1}^{2} - a_{2}^{2}} \\
\tan \bar{\phi} =& \frac{-a_{2}}{a_{1}} \,\,.
\end{align}
\end{subequations}


\subsection{Markov Chain Monte Carlo}

In this study we wish to characterize what we can learn about a wavelet present in the data. To accomplish this we marginalize the posterior distribution via the Markov Chain Monte Carlo (MCMC) algorithm. Suppose we inject a signal into our data $\textbf{s}$. Upon choosing a model specified by the initial set of parameters $\vec{x}$ we generate a proposed set of parameters from a probability density $q(\vec{y}|\vec{x})$. The chance that we accept this new set of parameters $\vec{y}$ is given by the Hasting's ratio

\begin{equation}
H = \min \bigg \lbrace 1, \frac{p(\textbf{s}|\vec{y},\mathcal{M}) p(\vec{y}|\mathcal{M}) q(\vec{x}|\vec{y})}{p(\textbf{s}|\vec{x},\mathcal{M}) p(\vec{x}|\mathcal{M}) q(\vec{y}|\vec{x})}  \bigg \rbrace \,\,.
\end{equation}
The sequence of parameters we accept, called a chain, constitute samples from the posterior distribution $p(\vec{\lambda}_{\tiny \mathcal{M}}|\textbf{s}, \mathcal{M})$. The MCMC we created used the F-statistic likelihood, extremizing the likelihood over the extrinsic parameters of the signal. This effectively reduces the search space of the MCMC, greatly improving its convergence, especially for the burst model which otherwise converges slowly when the sky location is poorly-constrained.

For the MCMC developed in this study uniform priors were set for the parameter set $\lbrace \log{A}$, $f_{0}$, $t_{0}$, $\log{\tau}$, $\bar{\phi}$, $\cos\theta$, $\phi$, $\psi$, $\epsilon \rbrace$. To aid in the convergence of the MCMC we used a mixture of proposal distributions. We utilized local Gaussian approximations to the posteriors through the Fisher matrix (which approximates the inverse covariance matrix)

\begin{equation}
\Gamma_{ij} = \sum_{\mathrm{I}} \left(\textbf{h}_{\mathrm{I},i}|\textbf{h}_{\mathrm{I},j}\right) \,\,,
\label{equ:Fisher}
\end{equation}
where $\textbf{h}_{I,i}$ represent the derivative of waveform (in the $I^\mathrm{th}$ data channel) with respect to the $i^{\mathrm{th}}$ parameter $\lambda_{i}$. These derivatives were calculated numerically using finite differencing of the waveforms discussed in section~\ref{sec:model}. We occasionally used proposals from the prior distribution. Since we are not currently developing a detection algorithm, only an MCMC which characterizes the signal, we used a targeting distribution to help the MCMC find appropriate central frequencies $f_{0}$ and decay factor $\tau$. For $f_{0}$ and $\tau$ individually, this proposal consisted of a Gaussian distribution centered on the true parameter used to generate the injection. The width of Gaussian was chosen based on the Fisher matrix estimation for the error in that parameter. To improve the acceptance rate of this proposal distribution the Gaussian was mixed with a 20\% by weight uniform distribution covering the prior range. Differential evolution~\cite{Braak2006} proposals were also used. Lastly, a time shift proposal was used to help highly oscillatory wavelets where shifts, forwards or backwards, in the central time of the wavelet by the wavelet's period were proposed (with an appropriate shift in initial phase).

To further improve convergence, and to ensure a thorough exploration of parameter space---such as investigating the existence of secondary modes on the sky for bursts---parallel tempering~\cite{PhysRevLett.57.2607} was utilized. During parallel tempering multiple chains are simulated simultaneously at different temperatures, i.e. their likelihoods are flattened $p(\textbf{s}|\vec{\lambda},\mathcal{M})^{\beta_{j}}$, where $\beta_{j} = 1/T_{j}$ is the inverse temperature fo the j$^{\mathrm{th}}$ chain. The cold chain, i.e. $T_{0} = 1$, represents samples from the posterior distribution. The chains at various temperatures propose and accept new parameters just as before, but with the flattened likelihood. Occasionally, swaps of parameters between chains neighboring in temperature are proposed based on the probability
\begin{equation}
H_{\mathrm{PT}} = \min \bigg \lbrace 1, \frac{  p(\textbf{s} | \vec{\lambda}_{j}, \mathcal{M})^{\beta_{j+1}}  p(\textbf{s} | \vec{\lambda}_{j+1}, \mathcal{M})^{\beta_{j}}   }{p(\textbf{s} | \vec{\lambda}_{j}, \mathcal{M})^{\beta_{j}}  p(\textbf{s} | \vec{\lambda}_{j+1}, \mathcal{M})^{\beta_{j+1}} }\bigg \rbrace \,\,.
\end{equation}
Parallel tempering vastly improves convergence once a proper selection of temperatures is made. The maximum temperature is chosen such that the hottest chain freely explores the parameters' prior volume, while not so hot as to be redundant in the prior space exploration as cooler chains. In section~\ref{sec:TI} we see how parallel tempering additionally aids us in determining whether a glitch (and which one) or a burst best explains the data.


\section{Parameter Estimation }\label{sec:PE}

\begin{figure*}[th]
	\centering	\includegraphics[width=3.5in]{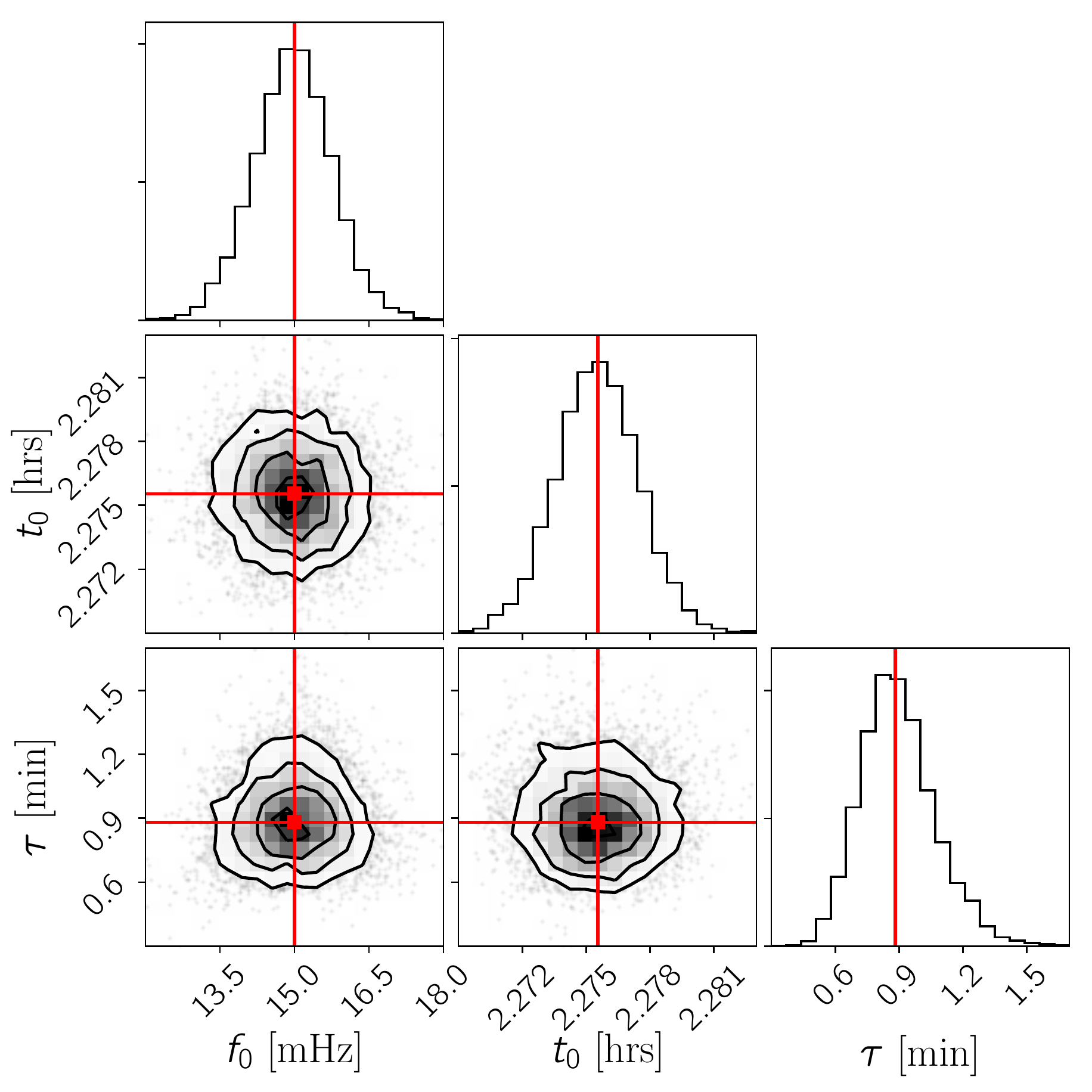} 	\includegraphics[width=3.5in]{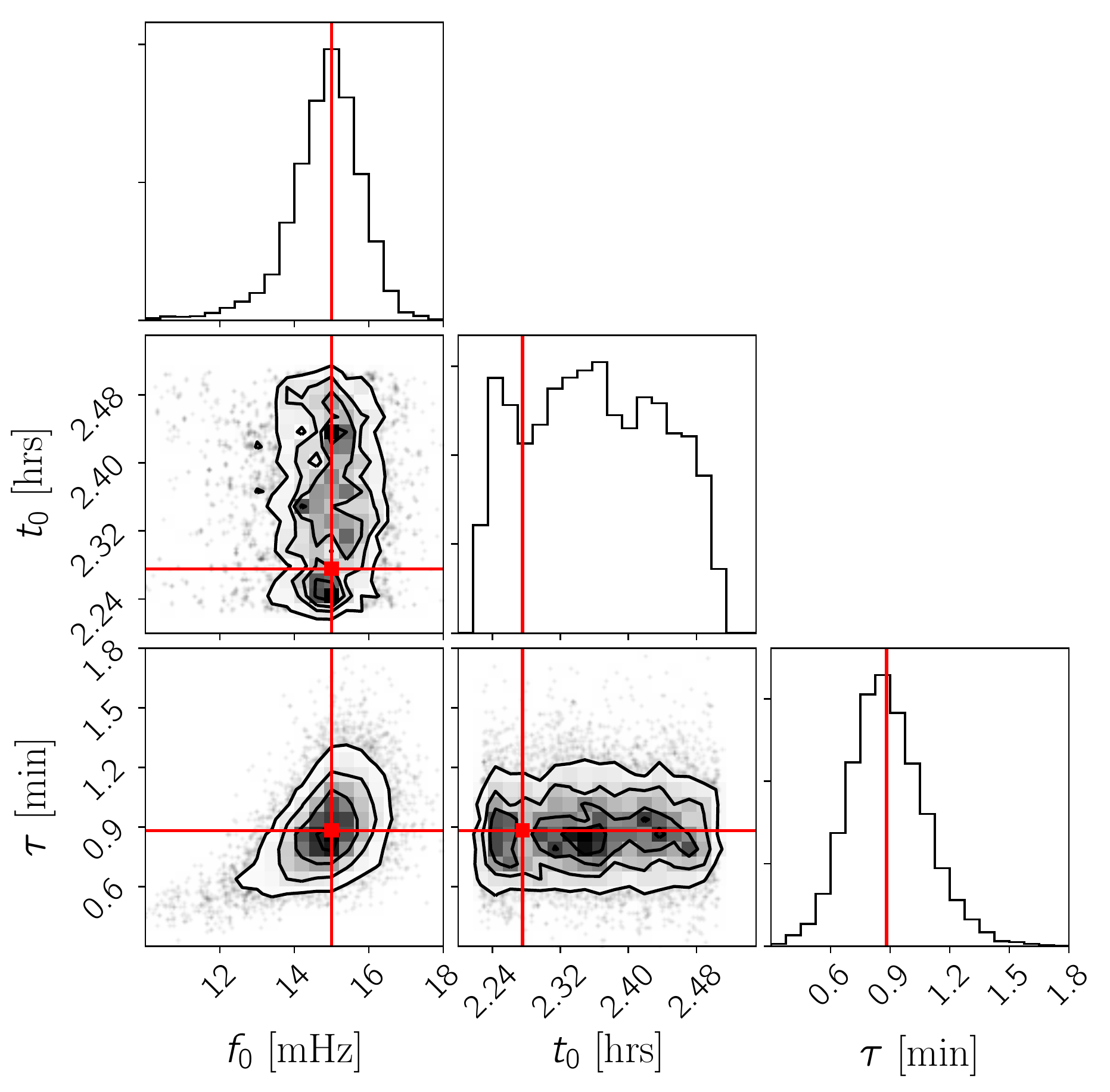}
	\caption{\label{fig:lowFreq_OP12} Marginalized posteriors for the parameters $f_{0}$, $t_{0}$, and $\tau$ are displayed for an optical path glitch $\Phi_{12}^{\mathrm{op}}$ on the left panel and a gravitational wave burst on the right panel. The fully marginalized posteriors for $f_{0}$ and $\tau$ are similar for the two injections (while the joint posterior exhibits some correlation for the burst), but the central time posteriors differ significantly. The injected parameters are marked by the red lines. }
\end{figure*}

The MCMC may now be used to address questions such as how well do we characterize the parameters of the signal, and how well do we recover the waveform itself. The central frequency $f_{0}$ and time damping factor $\tau$ are typically well determined for bursts and glitches. An example marginalized posterior for these parameters is seen in Figure~\ref{fig:lowFreq_OP12}. The left panel shows marginalized posterior distributions for an optical path glitch $\Phi_{12}^{\mathrm{op}}$ and the right panel shows marginalized posteriors for the same parameters for a burst. The injected signals both had a signal-to-noise of 8. The signal-to-noise (SNR) is given by

\begin{equation}
\rho = \sum_{I}(h_{I}|h_{I})\,\,.
\end{equation}
They shared the parameter values $f_{0} = 15$~mHz, $t_{0}=0.5 T$ (where $T$ is the observation period), and $\tau=53$~seconds (giving the wavelets a quality factor of 5.0). The burst injection had the following additional parameters: $\cos\theta=0.23$, $\phi=2.31$, $\psi = 0.45$, and $\epsilon = 0.5$. We see that the fully marginalized posterior distributions for the central frequency $f_{0}$ and $\tau$ are rather similar for these two injections. However, the posteriors for the central time $t_{0}$ are quite distinct in a significant way. One can show through a simple Fisher matrix calculation for a wavelet~\cite{0264-9381-32-13-135012} that the standard deviation in $t_{0}$ for a wavelet scales as $1/\rho\tau$. For the injected glitch has a measured standard deviation of $53$~seconds while the Fisher matrix standard deviation estimates an error of $70$~seconds, demonstrating agreement. The standard deviation in $t_{0}$ for the burst is $4.6$~minutes, much larger than that of glitch which must be attributed to the more complex response of a burst compared to a glitch. 

\begin{figure}[th]
	\centering	\includegraphics[scale=0.5]{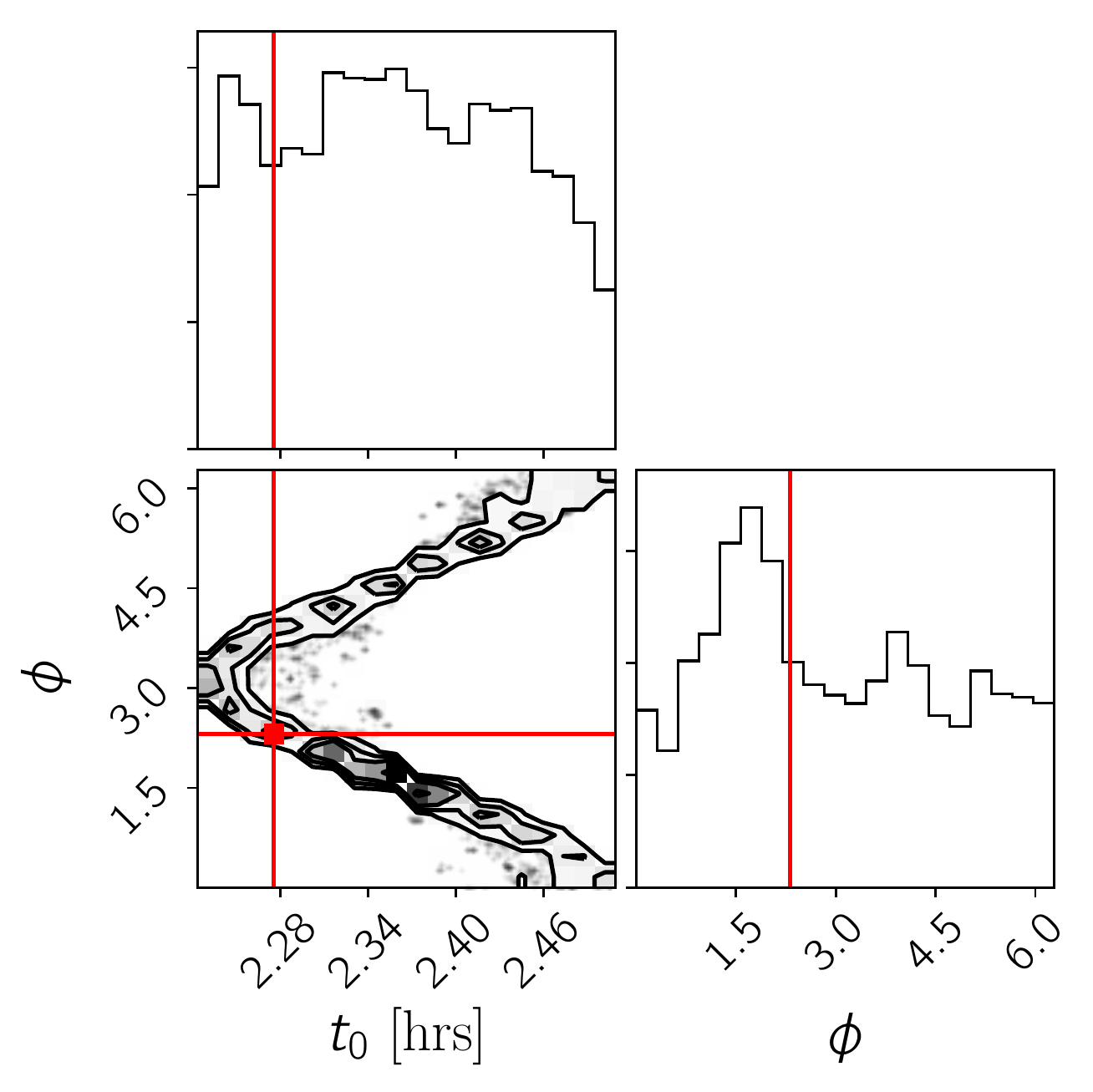}
	\caption{\label{fig:t0_phi} The figure displays the joint posterior for the azimuthal angle $\phi$ and the wavelet's central time $t_{0}$ as well as their fully marginalized posteriors. The injected parameters are again marked by the red lines. There exists a very large correlation between these two parameters.}
\end{figure}

The reason for the increase in error associated with the central time of the burst can be seen in Figure~\ref{fig:t0_phi}.
There exists a substantial correlation between the azimuthal sky angle $\phi$ and $t_{0}$. Without the central time constrained appropriately it turns out that we cannot determine the sky location, which is the case for this example burst. We can understand this by looking at the low frequency response to a GW burst signal. In this regime, the  Michelson-equivalent $A$ and $E$ TDI channels are proportional to $\frac{f}{f_{*}}\tilde{\Psi} e^{-2\pi i f \hat{\textbf{k}}\cdot\textbf{x}_{1}}$ modulo overall constants that differ between the channels. The $T$ channel is null in this limit. We see that the sky angles enter the phasing through time shift factor multiplying the Fourier transform of the Gabor-Morlet wavelet. Since this factor is a time shift, the sky angles are almost perfectly degenerate with the central time of the wavelet. The likelihood is approximately constant under mappings of the azimuth sky location and central time that keep fixed the combination $t_{0} - R \sin\theta\cos(2\pi f_{m} t_{0} - \phi) \,,$
where $f_{m} = 1/\mathrm{yr}$ is orbital modulation frequency and $R = 1$~AU. This relationship holds to leading order in the orbital eccentricity. Higher order corrections to the phasing incorporate additional information about the sky location in the form of the projected arm lengths $L_{ij} = L\left(1-\hat{\textbf{k}}\cdot\hat{\textbf{r}}_{ij}\right)$. 
\begin{figure}[th]
	\centering	\includegraphics[scale=0.4]{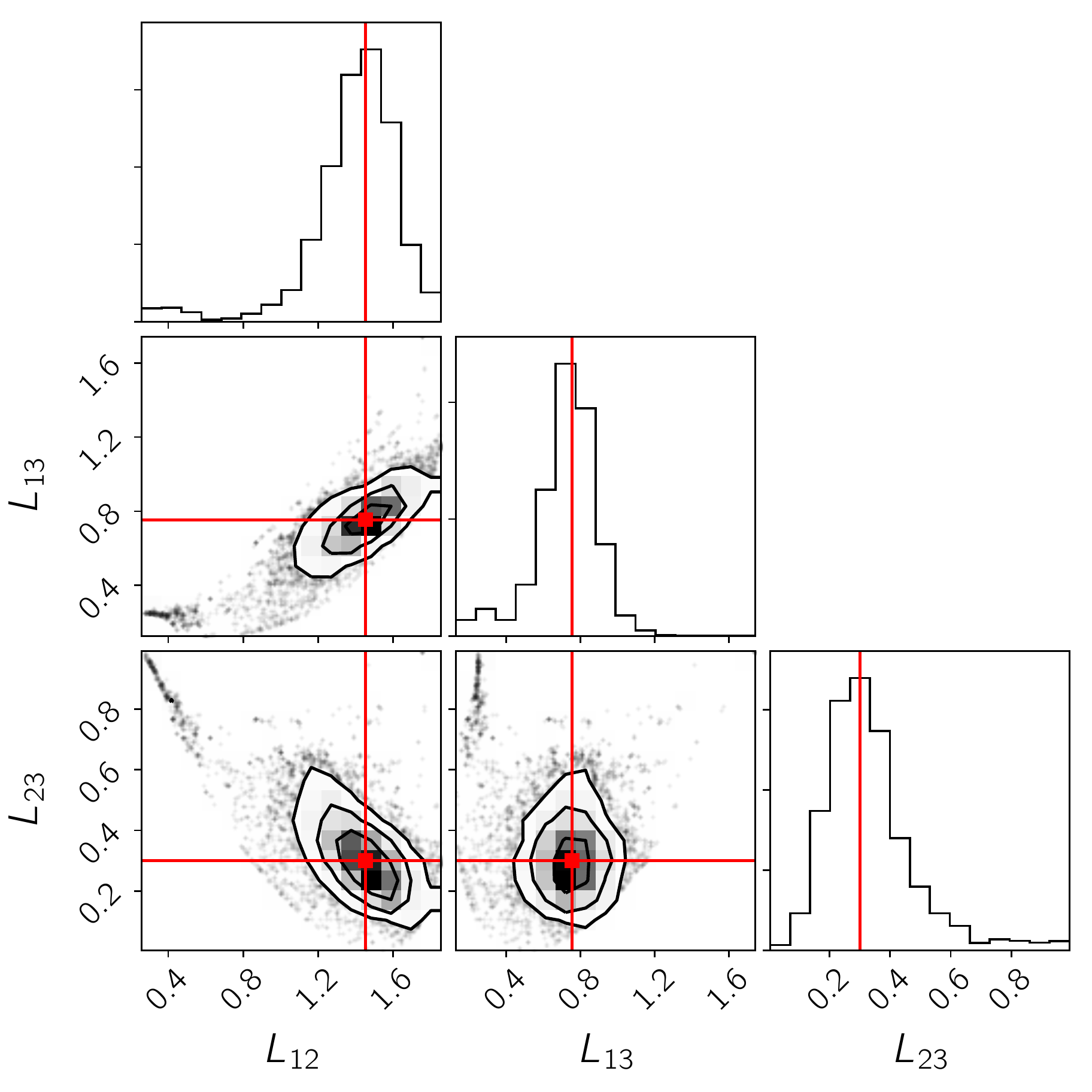}
	\caption{\label{fig:tDelay} The posterior distributions for the quantities \\$1-\hat{\textbf{k}}\cdot\hat{\textbf{r}}_{ij}$ are displayed here, i.e. LISA's projected arm lengths for a high frequency gravitational wave burst, normalized by the arm length $L=2.5$~Gm. The red lines indicate the injected values for the GW burst signal.}
\end{figure}

In Figure~\ref{fig:tDelay} we see the posteriors distribution for the projected arm lengths from a burst injection of the same sky location, polarization, ellipticity, and SNR as in the previous example, but now the central frequency is $50$~mHz and $\tau = 16$~seconds. This shorter envelope allows the central time to be measured and therefore the sky location to be better determined. The duration of the wavelet is more important in determining the sky location than the central frequency.

\begin{figure*}[ht]
	\centering	\includegraphics[width=3.5in]{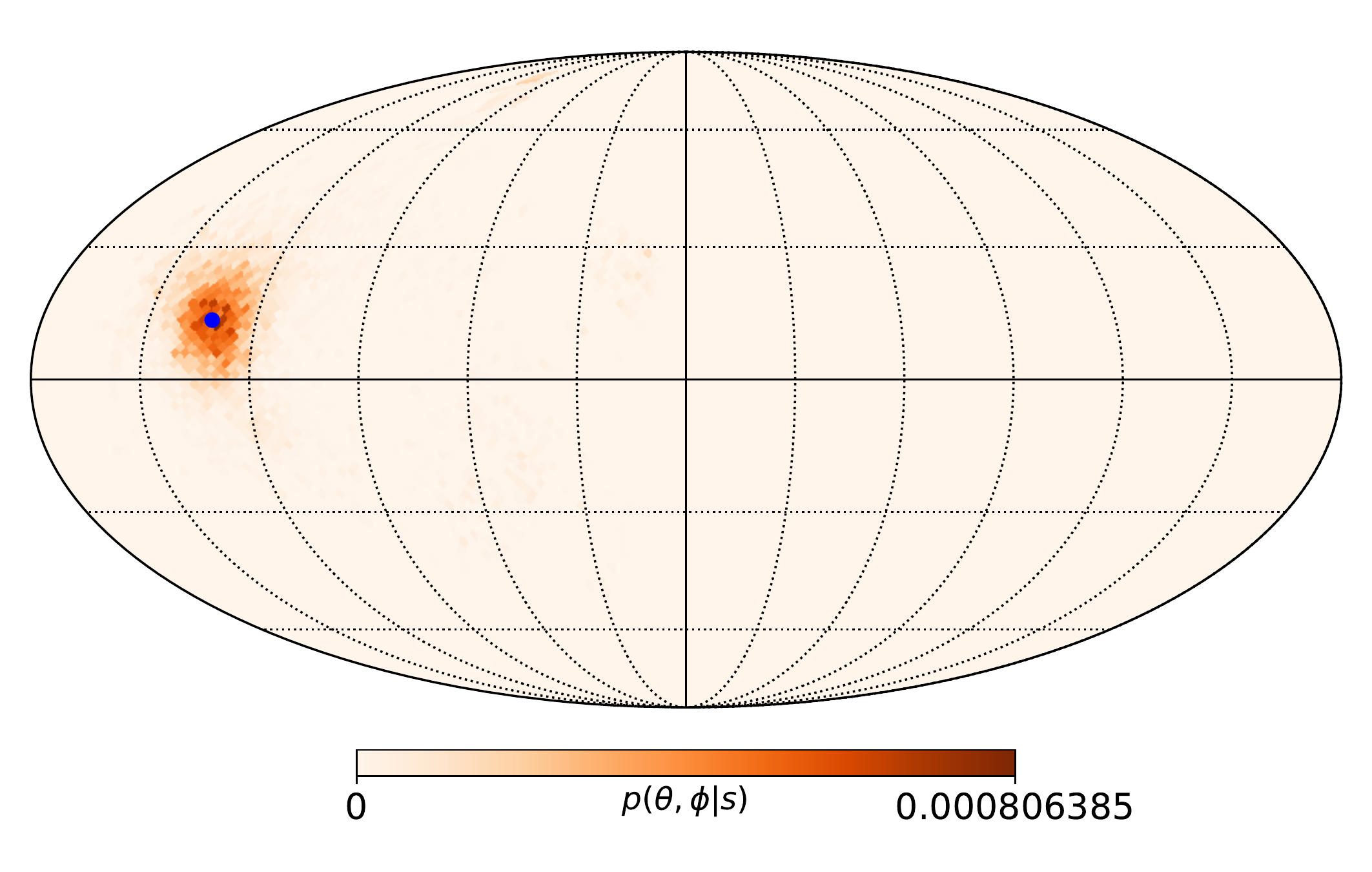} \includegraphics[width=3.5in]{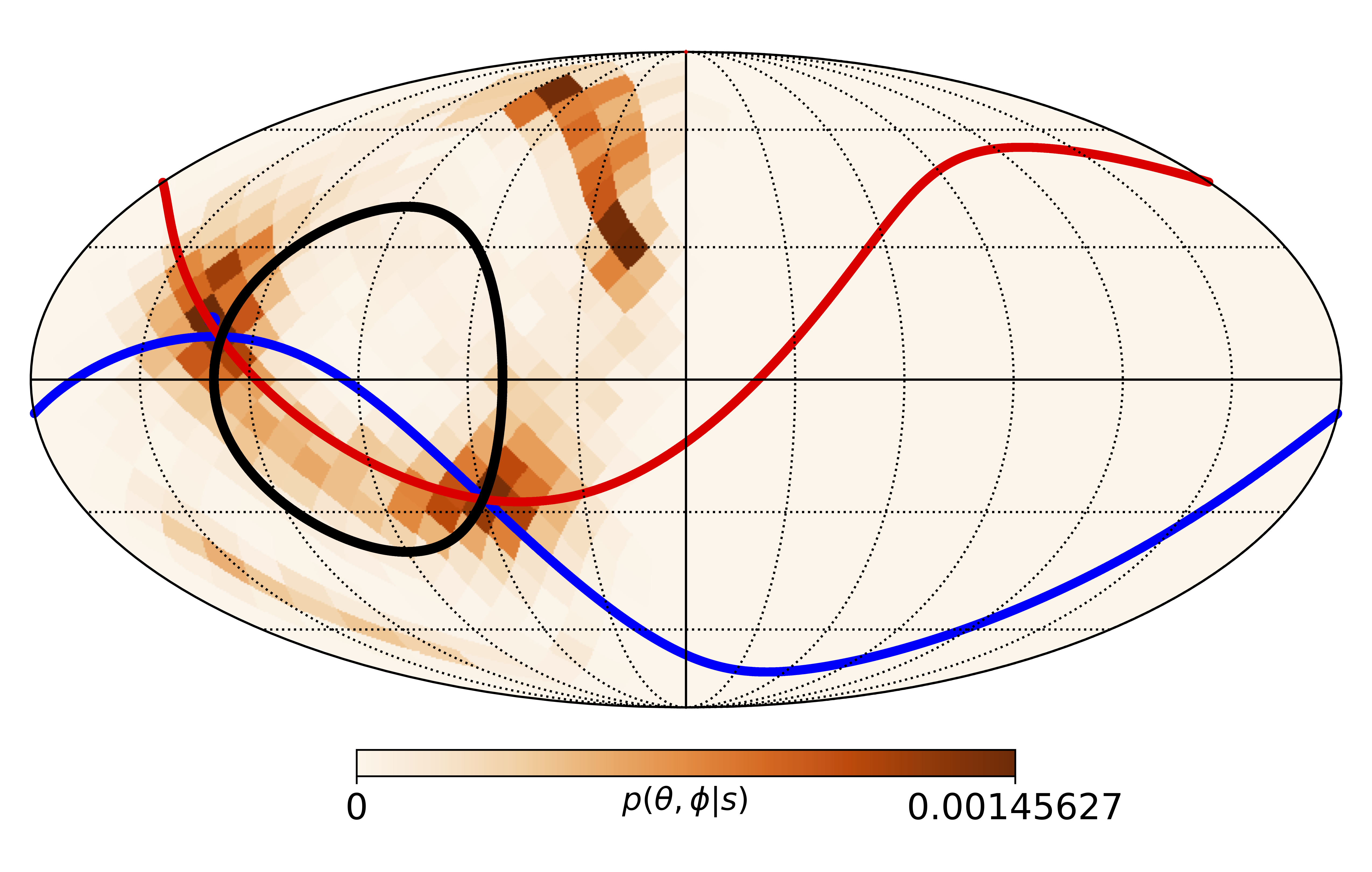} 
	\caption{\label{fig:sky}These figure displays the joint posterior for the sky location for two bursts of the same central frequency. The burst shown on the left has a short duration ($\tau = 17$~seconds), while the burst shown on the right has a longer duration ($\tau = 2.8$~minutes). The blue dot on the left figure represents the true sky location, which is the same for both sources. Lines of constant projected arm length are shown in the sky map on the right. The blue line for spacecraft $1$ and $2$, the red line for $1$ and $3$ and the black line for $2$ and $3$. There are two sky locations which satisfy these constraints. There is an additional maxima away from the intersection of these lines that corresponds to a secondary mode with an overall half-period time shift.}
\end{figure*}

In Figure~\ref{fig:sky} we see the posterior distribution for the sky location for two different high frequency bursts. Both bursts in Figure~\ref{fig:sky} have the same sky location, which is denoted by the blue dot in the sky map on the left. The central frequency of each source is $50$~mHz. The duration of the source shown on the left is $\tau=16$ seconds (i.e. the same burst used in Figure~\ref{fig:tDelay}) while the sky map on the right is for a source with $\tau=2.8$~minutes. We see that the origin of the burst on the sky has been localized for the short duration burst. This is due to the tight constraint on the central time of the wavelet. When the central time of the wavelet is measured to better than the light travel time between spacecraft we begin to have the power to localize the wavelet on the sky. The source shown on the right had a longer duration and poorer constraint on the central time, and therefore a poorer constraint on the sky location. Interesting structure emerges on the sky posterior for the right source. The most important factor in determining the sky location are the measured values for the projected arm lengths. The projected arm lengths can be related to the sky locations via the relations

\begin{widetext}
	\begin{subequations}
		\begin{align}
		\hat{\textbf{k}}\cdot\hat{\textbf{r}}_{12}(t) =& - \frac{1}{2}\cos\theta\left[\cos \alpha + \cos(\alpha + \frac{\pi}{3})\right] + \frac{1}{8 \sqrt{3}}\sin\theta \left[ 2 \cos(2\alpha - \phi) - 9 \cos\phi + 3\sqrt{3}\sin\phi + 2 \sin (\phi - 2 \alpha + \frac{\pi}{6})\right] \\
		\hat{\textbf{k}}\cdot\hat{\textbf{r}}_{13}(t) =& -\frac{1}{2} \cos\theta\left[ \cos\alpha + \sin(\alpha + \frac{\pi}{6})\right] + \frac{\sqrt{3}}{24}\sin\theta\left[2\cos(2\alpha - \phi) -9 \cos\phi -3\sqrt{3}\sin\phi+ 2\sin(2\alpha - \phi + \frac{\pi}{6}) \right]\\
		\hat{\textbf{k}}\cdot\hat{\textbf{r}}_{23}(t) =&-\frac{\sqrt{3}}{2}\cos\theta\sin\alpha + \sin\theta\left[ \sin(2\alpha - \phi + \frac{\pi}{6}) - 3\sqrt{3}\sin\phi + \sin(2\alpha - \phi - \frac{\pi}{6})\right] \,\,,
		\end{align}
		\label{eq:timeD}
	\end{subequations}
\end{widetext}
where $\alpha = 2 \pi f_{m} t$ . When these values, $\hat{\textbf{k}}\cdot\hat{\textbf{r}}_{ij}(t)$ which are present in the phase, are well measured then the sky location can be determined. In the right panel of Figure~\ref{fig:sky} we see the curves on the sky defined by equations~(\ref{eq:timeD}). The blue curve defines sky locations which give the same time delay (also same projected arm length) along the arm connecting the spacecrafts $1$ and $2$ specified by the true sky location of the injected burst. The red line displays the sky locations which maintain the same time delay between spacecraft $1$ and $3$ as the true sky location. The black line applies to the arm spanned by spacecraft $2$ and $3$. We see that these curves intersect at two specific sky locations, one of them coinciding with the true sky location by construction. The other intersection constituted a second mode that the MCMC explored. There exists one other mode which corresponds with a different central time that also provided a good fit to the data. This secondary mode is shifted by a half period in time from the true value. In addition to the time delays, the sky localization is also impacted by the antenna patterns which change the amplitude of the signal in each channel, but this is a weaker effect.

\begin{figure*}[th]
	\centering	\includegraphics[width=1.0\textwidth]{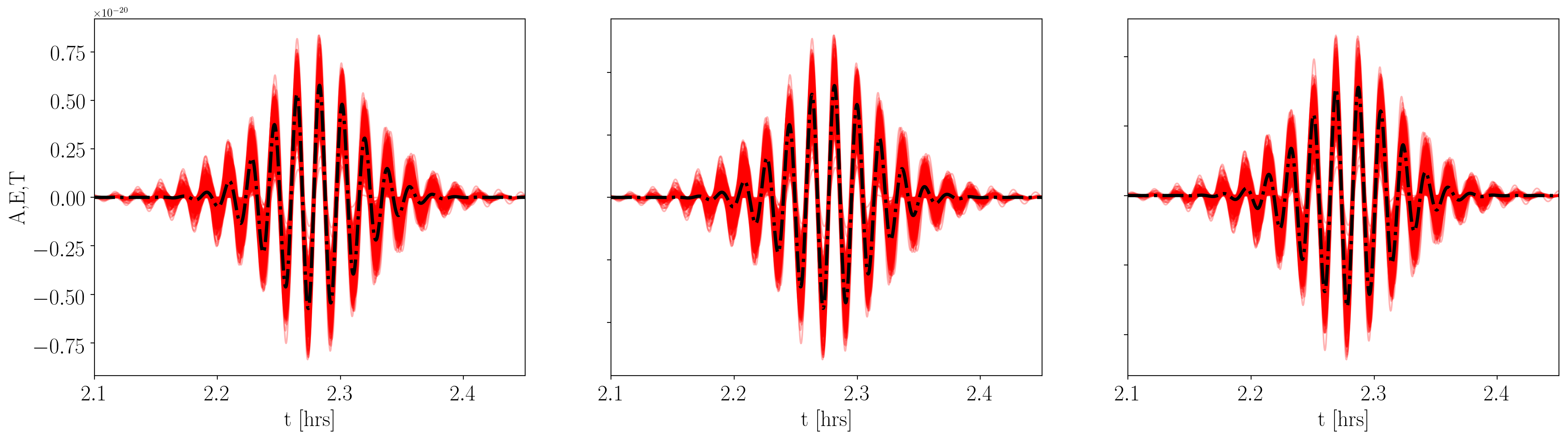}
	\caption{\label{fig:recOP12}  The left, center, and right panels display the $A E T$ TDI responses respectively for an optical path glitch injection denoted by the dot-dash black line. The red lines are MCMC samples of the waveform. The central frequency for this wavelet was $15$ mHz, and $\tau=2$~minutes.}
\end{figure*}

Lastly, we will also be concerned with the accuracy of our waveform reconstruction. Especially in future work where we will work with signals consisting of a superposition of wavelets. Figure~\ref{fig:recOP12} shows an example optical path glitch $\Phi_{12}^{\tiny \mathrm{op}}$ with a signal-to-noise of $8$, a central frequency of $15$ mHz, $\tau$ of $2$ minutes ($Q=17.0$). The observation period was set to $4.55$ hours. The dotted black line denotes the signal corresponding to the injected parameters. The red lines denote waveforms for parameters sampled from the MCMC.
Where the amplitude of the wavelet is largest we see that the MCMC sampled wavelets hug the injected waveform more tightly; the errors in the wavelets are greater near the edges.


\section{Model Selection }\label{sec:ModSel}

Now we will investigate whether we will be confused when identifying whether the data contains a glitch or signal. Recall that glitches entered the data stream with time delays of the arm light travel times. On the other hand, gravitational wave bursts instead enter the data stream with delays equal to the \textit{projected} arm lengths $L(1-\hat{\textbf{k}}~\cdot~\hat{\textbf{x}}_{ij})$. This impacts the phasing of there response, which can be used to infer the origin of glitch. Additionally, there are important differences between GW bursts and instrument glitches in where they place power in the TDI channels. Gravitational wave signals whose frequency is below the transfer frequency have a greatly diminished response in the $T$ channel. Additionally, the fact that gravitational waves are seen in at least two TDI channels, while glitches enter in only 1 or 2, will be of great importance. In this section we first demonstrate through a simple argument that we do not expect to be confused between GW signals and glitches when the all six laser links are operational. Later in this section we will more rigorously demonstrate this conclusion through calculation of the Bayesian evidence. We will also study whether it is the phasing of the response or the power distributed in the TDI data channels that provides the greatest leverage for separating GW bursts from instrument glitches.

If one has access to the $A$, $E$, $T$ data channels it is easy to make an argument that we will almost never confuse a glitch for a signal. Consider the following: noise in the data streams will affect our ability to match the true signal. A measurement of this match is the fitting factor (FF), i.e. a normalized (such that 1 indicates a perfect reconstruction of the signal)  noise-weighted inner product, between the data and model waveform, maximized over all model parameters. The noise in the data leads to statistical deviations in the fitting factor, even if the true parameters and model are used. The expected deviation from a fitting factor of 1 is described by~\cite{PhysRevLett.118.051101}

\begin{equation}
1- FF = \frac{D-1}{2\rho^2} \,\,,
\label{eq:FF_dev}
\end{equation}
where $D$ is the dimension of the model. 

Let us consider the scenario of using an acceleration glitch model, which crops up in the $X$ channel only, when the data actually contains a gravitational wave burst. In the low frequency limit, were we would expect to be most confused, the burst does not have significant power in the $T$ channel and also $A\sim h_{+}$, $E\sim h_{\times}$. Recall, in the frequency domain $\tilde{h}_{\times}  = i \epsilon \tilde{h}_{+}$. The overlap (normalized noise-weighted inner product) between the acceleration glitch and burst is

\begin{align}
&\frac{\sum_{I}(\textbf{s}_{I}|\textbf{h}_{I})}{\sqrt{\left(\sum_{I}(\textbf{s}_{I}|\textbf{s}_{I})\right)\left(\sum_{J}(\textbf{h}_{J}|\textbf{h}_{J})\right)}} = \nonumber \\ &~~~~~~~\frac{\frac{2}{3}(\textbf{A}_{B}|\textbf{X}_{\mathrm{ac}})}{\rho\sqrt{\left(\frac{2}{3}\right)^{2}(\textbf{X}_{\mathrm{ac}}|\textbf{X}_{\mathrm{ac}}) + \left(\frac{1}{3}\right)^{2}(\textbf{X}_{\mathrm{ac}}|\textbf{X}_{\mathrm{ac}})}} \,\,,
\end{align}
where $\rho$ is the SNR of the burst injection. This overlap is maximized if somehow the acceleration glitch conspires to be proportional to the burst's $A$ channel response $\textbf{X}_{\mathrm{ac}} \propto \textbf{A}_{\mathrm{B}}$. Let us also assume that the $A$ channel response to the burst accounts for a fraction $x$ of the squared SNR, i.e. $(\textbf{A}_{\mathrm{B}}|\textbf{A}_{\mathrm{B}}) = x \rho^{2}$. We then find that the overlap simplifies to $2\sqrt{x/5}$. If in the worst case scenario, all of the burst's SNR is in the $A$ channel the largest fitting factor that can be obtained is $0.89$---similar considerations for all other glitches demonstrate that this glitch is indeed the worst case in the regime under consideration.

Are we to be concerned by a fitting factor this large? To answer this question we can consider eqn.~($\ref{eq:FF_dev}$) to understand the statistical error in the fitting factor. Inserting the value $0.89$ into this equation results in an SNR of $4.3$. We can loosely understand this SNR as the largest SNR possible for the burst that could result in a confusion of the origin of the data (i.e. whether it was a glitch or signal). So we see that, under some very general assumptions, it is only when a burst is marginally detectable that we might confuse it for an instrument glitch.


\subsection{Bayesian Evidence} \label{sec:TI}

To more rigorously find out which model best explains the data we must calculate the ratio of evidences $p(\textbf{s}|\mathcal{M})$ for two given models. This quantity is known as the Bayes factor

\begin{equation}
B_{ij} = \frac{p(\textbf{s}|\mathcal{M}_{i})}{p(\textbf{s}|\mathcal{M}_{j})} \,\,.
\end{equation}
In this subsection we calculate the Bayes factor for competing glitch and burst models for different injections such that we can understand our ability to distinguish a signal's origin and to determine when we cannot. We calculate the Bayes factor via thermodynamic integration~\cite{PhysRevD.80.063007}. Since we have utilized parallel tempering in our MCMC we can calculate the average log likelihood $E_{\beta}\left[ \log p(\textbf{s}|\vec{\lambda},\mathcal{M})\right]$ for each temperature of the MCMC by simply calculating the sample mean of the log likelihood values for each sample in the chain. With these in hand one may calculate the evidence for a model via the integral

\begin{equation}
\ln p(\textbf{s}|\mathcal{M}) = \int_{0}^{1} d\beta E_{\beta}\left[ \log p(\textbf{s}|\vec{\lambda},\mathcal{M})\right] \,\,.
\end{equation}
We perform the integral using the methods described in Refs.~\cite{PhysRevD.91.084034,0264-9381-32-13-135012}. The covariance matrix between the log likelihood values for each temperature are estimated and used to define a log likelihood for the thermodynamic integration integrand.~\cite{PhysRevD.91.084034}. The integrand is fit by a cubic spline whose control points and locations are marginalized over via a reversible jump MCMC~\cite{Green1995}. The MCMC gives us estimates for the evidence integral (upon integrating the cubic spline) and its associated error.

\begin{figure}[th]
	\centering	\includegraphics[scale=0.45]{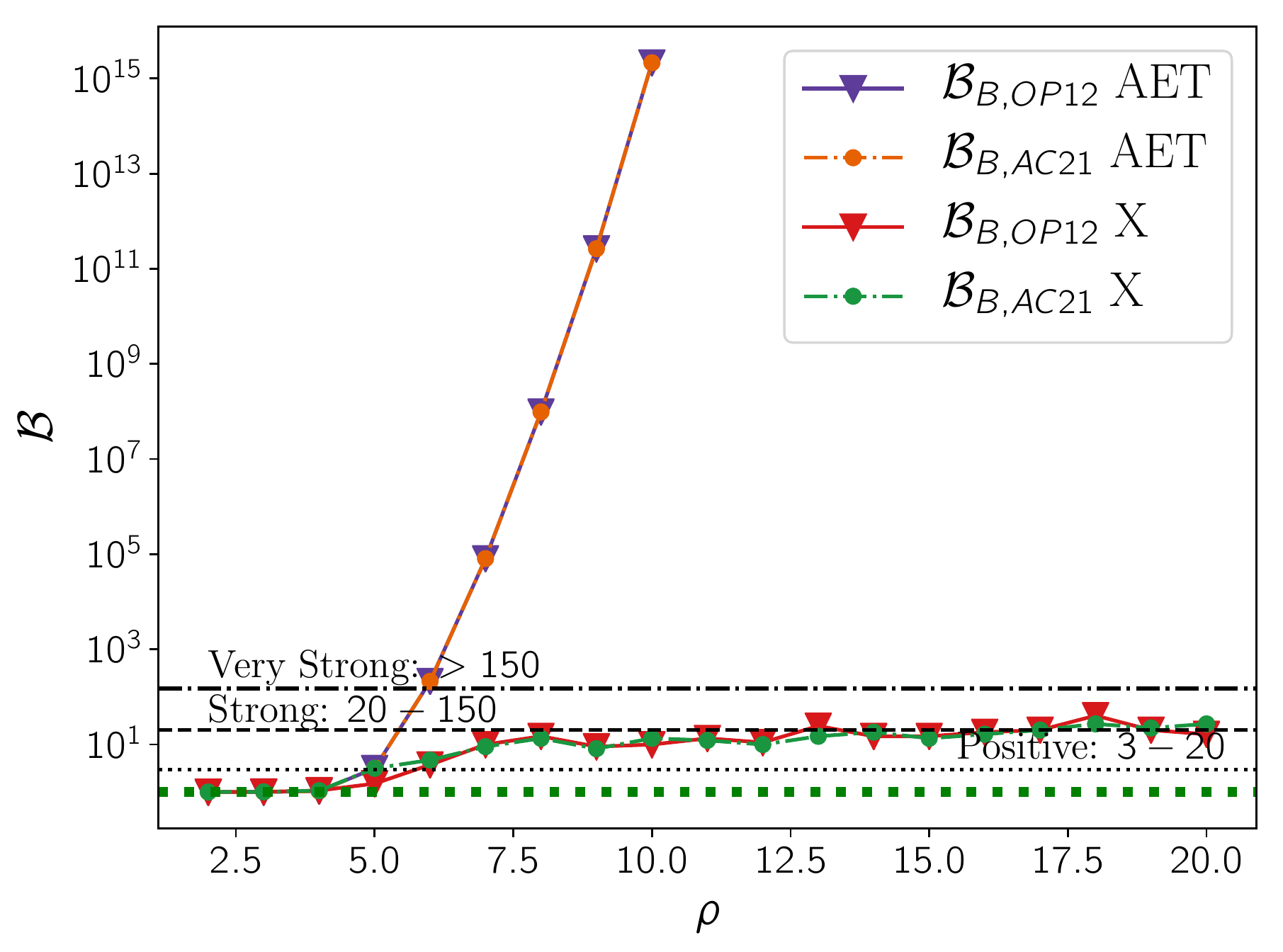}
	\caption{\label{fig:BF_burst1} Bayes factors as function of signal to noise ratio. A burst was injected into the data stream with a central frequency of $15$~mHz and $\tau =2$~minutes. The purple and orange lines represent the Bayes factors when the $AET$ TDI channels were used. Note that for the $AET$ lines the Bayes factor lie on top of each other. The red and green lines represent the case when only the $X$ TDI channel was used. Triangle markers denote the GW burst Bayes factors vs $\Phi_{12}^{\mathrm{op}}$, while circle markers represent GW burst Bayes factors vs $\Phi_{21}^{\mathrm{ac}}$. With the $AET$ channels we gain confidence swiftly of the true model as SNR grows. The growth in Bayes factor is much slower when only the $X$ channel is available.}
\end{figure}
\begin{figure}[th]
	\centering	\includegraphics[scale=0.45]{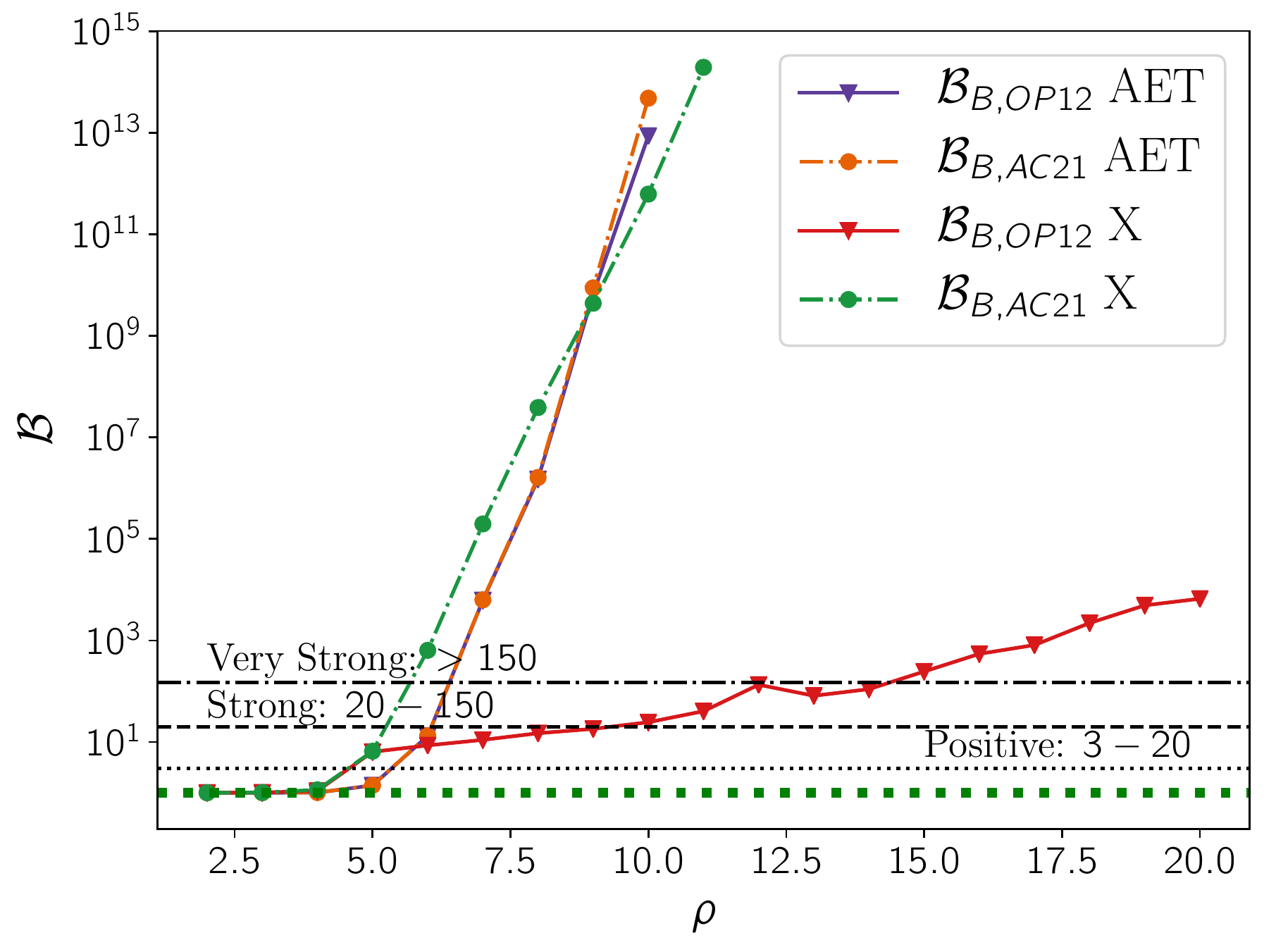}
	\caption{\label{fig:BF_burst2} Bayes factors as function of signal to noise ratio for a high frequency burst injection. The lines are labeled according to the same scheme as in Figure~\ref{fig:BF_burst1}. The GW burst is much easier to separate from an instrument glitch in this case.}
\end{figure}

Figure~\ref{fig:BF_burst1} shows the Bayes factor between the glitch and GW burst models for data containing a simulated burst. We use the notation $\mathcal{B}_{A,B} = p(\textbf{s}|A)/p(\textbf{s}|B)$ to represent the Bayes factor in the figure legends. Additionally, the label $B$ is used to denote the burst model. The burst injections has the same values for $\theta$, $\phi$, $\psi$, and $\epsilon$ as the burst discussed in section~\ref{sec:PE}. The other important parameters are the central frequency, set to $15$~mHz, and $\tau$ set to $2$~minutes (giving a quality factor of $17.0$). The orange and purple lines denote the Bayes factor when the $AET$ TDI channels are used and the red and green lines show the Bayes factor when only the $X$ channel was used. The lines marked by upside-down triangles represent the model comparison between the burst model and an optical path glitch between spacecraft $1$ and $2$. The lines marked with circles represent the Bayes factor between the burst model and an acceleration glitch between the two same spacecraft. The dashed green line represents a Bayes factor of 1, {\it i.e.} no preference between the two competing models. A Bayes factor between $3-20$ shows positive evidence~\cite{Romano2017} towards the true model. The evidence of the correct model is strong if the Bayes factor lies in the range $20-150$, and considered very strong if the Bayes factor is greater than 150. These regions are denoted by the various dashed horizontal black lines in figures~\ref{fig:BF_burst1} and~\ref{fig:BF_burst2}. With the $AET$ channel combination we see that the Bayes factor grows rapidly with signal-to-noise, and for SNRs greater than 5 we are confident that the signal is astrophysical. This supports our argument that GW burst and glitches are easily separated when we have the full collection of TDI channels. With just the $X$ channel the prospects are not as good, and it is not until the signal reaches SNR 10 that it can be confidently distinguished from a glitch. In this low frequency regime we find that a burst injection recovered with an optical path glitch model gets biased central frequency and damping time scales. In Figure~\ref{fig:BF_burst2} we see the Bayes factors for a high frequency burst injection, where $f_{0} = 50$~mHz and $\tau=16.9$~seconds ($Q=5.0$). We see that it is much easier to differentiate a gravitational wave burst from an acceleration glitch. Our ability to distinguish this burst from the optical path glitch $\Phi_{12}^{\mathrm{op}}$ however is not enhanced as much as that of the acceleration glitch, but is still improved. 

\begin{figure}[th]
	\centering	\includegraphics[scale=0.45]{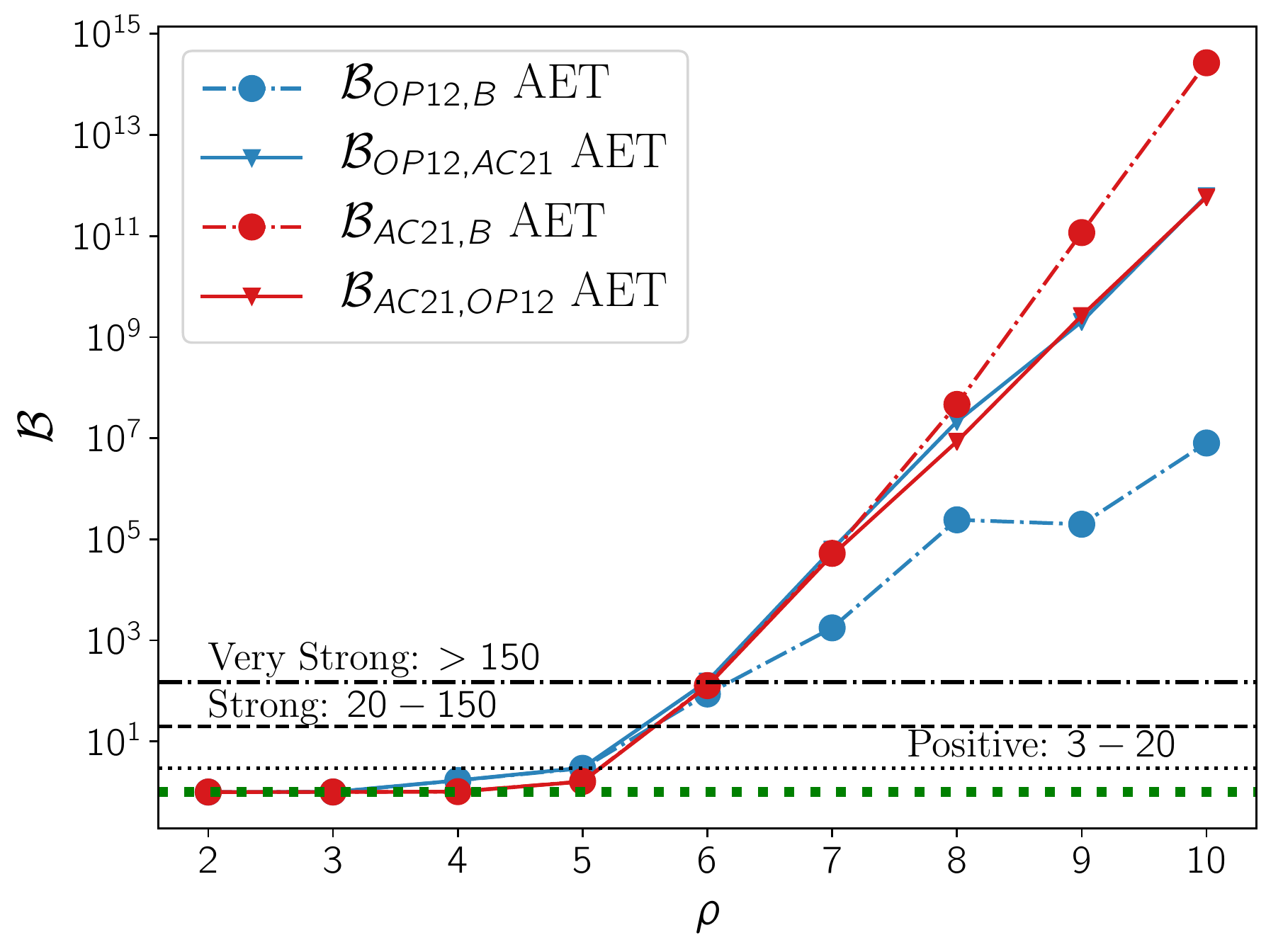} 
	\caption{\label{fig:glitchBayes} This figure displays Bayes factors for a high frequency glitch injections. The blue lines denote Bayes factors for an optical path glitch $\Phi_{12}^{\mathrm{op}}$ injection and the red lines denote a $\Phi_{21}^{\mathrm{ac}}$ injection. }
\end{figure}

Lastly, we wish to how well we can differentiate models for glitch injections. For low frequency injections the story is similar in that with the full $AET$ data stream we will be able to differentiate a glitch, both acceleration and optical path, through the distribution of power in the different data channels. When we only have the $X$ channel discrimination once again becomes challenging until the SNR becomes large. In Figure~\ref{fig:glitchBayes} Bayes factors are displayed for high frequency glitch injections. The central frequency of the glitches was $50$~mHz and $\tau=11$~seconds ($Q = 3.5)$. The blue lines represent Bayes factors for an optical path glitch injection $\Phi_{12}^{\mathrm{op}}$ and the red lines an acceleration glitch injection $\Phi_{21}^{\mathrm{ac}}$. We see that in this high frequency regime there will little issue in discriminating the origin of the signal. This figure also suggests, as seen before, that there might be a more of a challenging discriminating this optical path glitch from a  burst.

\section{Discussion }\label{sec:discuss}

To realize the full discovery potential of the LISA observatory we need to be in a position to detect unexpected and unknown signals. We have developed a forward model for a wavelet basis to represent instrumental glitches and gravitational wave bursts, as first step towards this goal. Ideally, to separate unmodeled signals from noise, we would have multiple independent LISA observatories. We have shown that the separation is possible with a single LISA detector, and even with a single TDI data channel, though the performance is much better when all three TDI channels are available. The properties of the signals and glitches can be recovered with good accuracy, though degeneracies in some parameters can degrade sky localization.

There are several extensions to that will need to be made to handle generic glitches and signals . In our analysis we assumed the Gaussian noise level were not only equal, but also that they were known. In reality power spectral density of the noise in each component will have to be estimated from the data, as was done in Refs.~\cite{PhysRevD.82.022002,PhysRevD.89.022001}. We will also need to generalize the analysis to model non-stationary noise, a complication we know LISA will experience owed at least in part to the significant contribution to the noise by unresolved galactic binaries~\cite{1742-6596-840-1-012024}. Our analysis in this paper took a quasi-Bayesian approach via the maximization of the likelihood over extrinsic parameters through the F-statistic. In the future a full marginalization will have to be done, though the the F-statistic could be used to produced very effective proposal distributions for the MCMC based on maps of the F-statistic likelihood.

For gravitational wave bursts we will generalize the polarizations to be not elliptically polarized. One last crucial extension is the use of multiple wavelets in the analysis~\cite{0264-9381-32-13-135012}. Not only will we need to characterize multiple wavelets, but we will also need to marginalize over the number of wavelets in the data stream. Due to the shear number of combinations of wavelets we expect in data stream we would expect the need to implement an effective Reversible Jump MCMC~\cite{Green1995} implementation to address the issue of determining an appropriate number of wavelets and for determining the evidence that a GW signal or an instrument glitch is present in the data. There may be additional information gather by LISA in the form of instrument monitors. These could provide crucial information in characterizing glitches and assessing whether a glitch has indeed occurred.

\section*{Acknowledgments}
TR and NJC appreciate the support of the NASA grant NNX16AB98G.

%
%

\bibliography{refs}

\end{document}